\documentclass[a4paper,fleqn,usenatbib]{mnras}
\usepackage{aasmacros}
\usepackage{times}
\usepackage{amsmath}
\usepackage{graphicx}
\usepackage{amssymb}
\usepackage[usenames,dvipsnames]{color}
\usepackage{url}
\usepackage{fixltx2e}
\usepackage{mathtools}
\usepackage{amsmath}
\usepackage{float}
\usepackage{cite}
\usepackage{fixltx2e}
\bibpunct{(}{)}{;}{a}{,}{,}
\usepackage{longtable}
\usepackage{hyperref}
\usepackage{morefloats}
\usepackage[T1]{fontenc}
\usepackage{ae,aecompl}
\usepackage{mathptmx}
\usepackage{txfonts}

\let\cite=\citen

\def\apj{ApJ}
\def\aj{AJ}
\def\apjs{ApJl}
\def\apjl{ApJl}
\def\mnras{MNRAS}
\def\pasa{PASA}

\def\pasp{PASP}

\def\aspcs{ASPCS}

\defcitealias{Condon12}{CO12}
\bibpunct{(}{)}{;}{a}{}{,}

\def\speak{$S_{\rm peak}$}
\def\thetamaj{$\theta_{\rm maj}$}
\def\thetamin{$\theta_{\rm min}$}
\def\posang{$\phi$}
\def\thetabeam{$\theta_{\rm B}$}
\def\stot{{$S_{\rm total}$}}
\def\thetadmaj{$\theta_{\rm Dmaj}$}
\def\thetadmin{$\theta_{\rm Dmin}$}
\def\arcs{\prime \prime}
\def\Ob{\textsc{Obit}}
\def\Ag{\textsc{Aegean}}

\makeatletter
\newcommand*{\rom}[1]{\expandafter\@slowromancap\romannumeral #1@}
\makeatother

\title[3-GHz catalogue simulations]{Deep 3-GHz Observations of the Lockman Hole North with the Very Large Array -- \rom{1}. Source extraction and uncertainty analysis}

\author[Vernstrom et. al]{T. Vernstrom\thanks{E-mail:vernstrom@dunlap.utoronto.ca}$^1$, Douglas Scott$^2$, J.V. Wall$^2$, J.J. Condon$^3$, W.D. Cotton$^3$,\newauthor
R.A. Perley $^4$ \\
  $^1$Dunlap Institute for Astronomy and Astrophysics, University of Toronto, Toronto, ON M5S 3H4, Canada\\
  $^2$Department of Physics and Astronomy, University of British Columbia, Vancouver, BC V6T 1Z1, Canada\\
  $^3$National Radio Astronomy Observatory, 520 Edgemont Rd, Charlottesville, VA 22903, USA \\
  $^4$National Radio Astronomy Observatory, P.O. Box 0, Soccoro, NM 87801, USA\\
}
\begin{document}

\pagerange{\pageref{firstpage}--\pageref{lastpage}} \pubyear{2016}

\maketitle

\label{firstpage}
\begin{abstract}
This is the first of two papers describing the observations and cataloguing of deep $3$-GHz observations of the Lockman Hole North using the Karl G. Jansky Very Large Array. The aim of this paper is to investigate, through the use of simulated images, the uncertainties and accuracy of source-finding routines, as well as to quantify systematic effects due to resolution, such as source confusion and source size. While these effects are not new, this work is intended as a particular case study that can be scaled and translated to other surveys. We use the simulations to derive uncertainties in the fitted parameters, as well as bias corrections for the actual catalogue (presented in Paper \rom{2}). We compare two different source-finding routines, \textsc{OBIT} and \textsc{AEGEAN}, and two different effective resolutions, 8 and $2.75\,$ arcsec. We find that the two routines perform comparably well, with \textsc{OBIT} being slightly better at de-blending sources, but slightly worse at fitting resolved sources. We show that 30 to 70 per cent of sources are missed or fit inaccurately once the source size becomes larger than the beam, possibly explaining source count errors in high-resolution surveys. We also investigate the effect of blending, finding that any sources with separations smaller than the beam size are fit as single sources. We show that the use of machine-learning techniques can correctly identify blended sources up to 90 per cent of the time, and prior-driven fitting can lead to a 70 per cent improvement in the number of de-blended sources.

\end{abstract}

\begin{keywords}

cosmology: observations -- radio continuum: galaxies --methods:data analysis -- methods: statistical

\end{keywords}

\section{Introduction}
\label{sec:introduction}
The making of a source catalogue from survey data can be a tricky business. Telescopes perform imaging of the sky, which we often want to compress into lists of sources and properties \citep[see e.g.][]{Hogg11}; however, there is no unique way to perform the cataloguing process, and many detailed choices need to be made along the way. There are different software packages available for source finding and extraction, each with advantages and disadvantages, and each generally aimed at specific types of data. The type of data being catalogued, i.e. wavelength, telescope type, imaging process, instrumental noise, resolution, etc, can introduce specific systematic uncertainties that have to be carefully treated. With newer radio interferometry data, details such as {\it uv}-coverage, wide bandwidth, and wide fields also play a role in the image-making process and ultimately the characteristics of the images being turned into catalogues. 

The next generation of radio surveys, such as a new deep survey with the Karl G. Jansky Very Large Array (VLA) \citep{Jarvis14}, the Evolutionary Map of the Universe \citep[EMU,][]{Norris11} with the Australian Square Kilometre Array Pathfinder (ASKAP), and the MeerKAT International GigaHertz Tiered Extragalactic Exploration survey \citep[MIGHTEE,][]{Jarvis11}, will all survey large areas of the sky with rms values ranging from $0.1$ to $10\, \mu$Jy beam$^{-1}$. Not only will these be the deepest and largest radio surveys to date, they will also produce more data than previous surveys. Data processing will need to be fully automated, including the source extraction procedure. Thus it is important to investigate source-extraction methods, uncertainties, corrections, and potential problem areas now for smaller data sets, in the hope of optimizing the process before these surveys commence. 

The majority of standard source-finding and extraction algorithms have been tested and shown to be highly reliable and complete \citep[e.g.][]{Hopkins15,Hancock12}, except for the faintest sources, usually near the detection limit. However, it tends to be these sources that we are most interested in, as telescope improvements allow us to reach new depths and thus probe fainter sources. 

We have known qualitatively since the 1960s and 70s about the difficulties in making reliable and complete catalogues due to the effects of noise, source size, blending, and the resulting effects on source counts and catalogues \citep[see e.g. ][]{Large61,Hill62,Wall71,Bridle72}. However, it is necessary to quantitatively address these issues taking into account modern telescopes and techniques. 

There have been several recent analyses aimed at investigating source finding in radio astronomy \citep[e.g.][]{Popping12,Huynh12,Mooley13,Hopkins15}. In this paper we present a detailed look at the cataloguing process for simulations of a single pointing, with specific conditions set to match our observations. Our goal here is first of all to derive a cataloguing process and the corrections to be used with our VLA observations, and secondly to provide additional quantitive information on a number of effects that can be used by future surveys.

We have obtained deep (instrumental rms $\sigma\simeq 1\, \mu$Jy beam$^{-1}$) $3\,$GHz data from the VLA \citep{Condon12} in the Lockman Hole North. Using simulations set up to mimic our VLA images we investigate different source-extraction algorithms and the effect of resolution on source fitting, as well as issues such as completeness, source blending, and source size. We note that this paper does not address any non-Gaussian noise effects or image artefacts arising from issues originating with the {\it uv} data. Our simulations and analysis are performed solely in the image plane. The catalogue obtained from the VLA data and details of the sources are presented in Paper \rom{2} \citep{Vernstrom16b}. 

The outline for this paper is as follows. In Section~\ref{sec:obs} we briefly describe the data obtained from the VLA and the image characteristics. Section~\ref{sec:cat_sims} describes the general process of source finding and fitting and the simulation setups. In Section~\ref{sec:cat_simerr} we discuss the results of the simulations in terms of parameter uncertainties and corrections. Section~\ref{sec:cat_cf} describes how we use the simulation results to look at completeness levels and false detection rates. In Sections~\ref{sec:sizes} and \ref{sec:blend} we detail the simulations and present results investigating the effects of source size and source blending on the fit results. Section~\ref{sec:disc} discusses our findings and compares the results from the different resolutions and software routines.

\section{Data}
\label{sec:obs}

The simulations and tests performed in this paper are designed to investigate the catalogue procedure for the data presented in Paper \rom{2}. These observations were made with the $S$-band receivers of the VLA , which span 2 to $4\,$GHz. The details of the observations (calibration, editing, imaging, etc) are described in more detail in \citet{Condon12}, \citet{Vernstrom13} and Paper \rom{2}. The VLA  was used to target a region in the Lockman Hole North, centred on $\alpha = 10^{\rm{h}}46^{\rm{m}}00^{\rm{s}}$, $\delta = +59^{\circ}01'00''$ (J2000). 

The observations were made with two VLA configurations, the C configuration and the BnA configuration (consisting of roughly 50 hours and 20 hours, respectively). We consider two separate data sets here: the C configuration data only (herein ``C data" or ``C image") and the BnA and C configurations combined (herein ``CB data" or ``CB image"). The C image has a circular $8\,$arcsec synthesized beam, while the CB image has a circular $2.75\,$arcsec synthesized beam. The instrumental noise in each image, before primary beam corrections, is $\sigma_{\rm n}=1.08 \,\mu{\rm Jy\,beam}^{-1}$ for the C data and $\sigma_{\rm n} = 1.15 \,\mu{\rm Jy\,beam}^{-1}$ for the CB data. A summary of these details is provided in Table~\ref{tab:image}. 

It is important to investigate the effect of resolution on catalogue results, in order to see how this might impact future surveys. A deep VLA  survey \citep[like that described in][]{Jarvis14} would have a resolution closer to our CB data (or better), whereas the EMU survey will have a lower resolution of $8$ to $10\,$arcsec, similar to our C data; the MIGHTEE survey has a range of resolutions from 3 to $8\,$arcsec depending on which tier is considered. Cataloguing the two resolutions of the same field, calibrated and transformed to images in the same way, should tell us about the differences that can be expected between the surveys. Since the images have comparable noise levels, cataloguing them separately allows us to investigate the effects of resolution and confusion on issues like uncertainty in the fits, completeness, and source blending. 

\begin{figure*}
\centering
\includegraphics[scale=.48,natwidth=15in,natheight=3.75in]{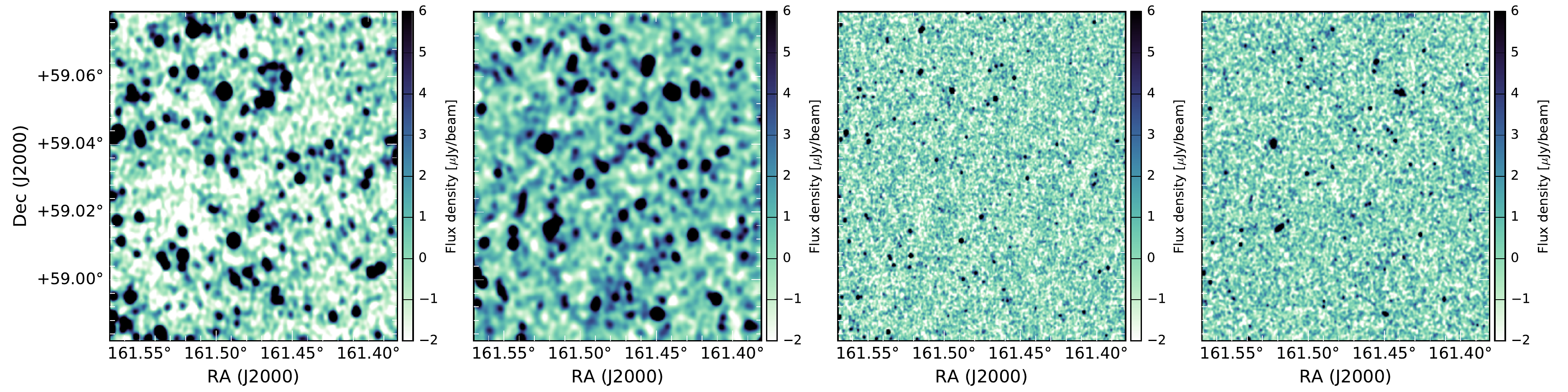}
\caption{Actual and simulated VLA images. The left panel shows a region of the actual C image, while the second panel shows the same sized region for one of the simulated C images. The third panel shows the same region of the actual CB image, while the rightmost panel is a region from a simulated CB image.}
\label{fig:simims}
\end{figure*}

\begin{table}
\centering
\caption{Primary beam-corrected image properties for the wide-band VLA data. The reported noise values are all after correction for the primary beam and frequency weighting effects, with $\rho$ being the distance from the pointing centre. The clean beam size, \thetabeam, is given as the full width at half maximum (FWHM), and the field of view (FOV) is the FWHM of the primary beam at $3\,$GHz.   }
\begin{tabular}{llll}
\hline
\hline
Quantity & C-data & CB-data& Unit \\
\hline
$\langle\nu\rangle (\rho=0)$ & 3.06 & 3.036&GHz\\
$\langle\nu\rangle(\rho=5^\prime)$ & 2.96 &2.96& GHz\\
Pixel size & 0.5 & 0.5&arcsec\\
\thetabeam & 8.00 &2.75& arcsec\\
FOV & 13.75 &13.75&arcmin \\
$\sigma_{\rm n}(\rho$=0) & 1.08 &1.15& $\mu$Jy beam$^{-1}$ \\
$\sigma_{\rm{n}}(\rho=5^\prime$)& 1.447 &1.54& $\mu$Jy beam$^{-1}$ \\
\hline
\end{tabular}
\label{tab:image}
\end{table}

\section{Setup and Simulations}
\label{sec:cat_sims}

The tests in this paper were run using two different source finding programs: the {\Ob} \citep{Cotton08}\footnote{\url{http://www.cv.nrao.edu/~bcotton/Obit.html}} task \textsc{FndSou}; and the {\Ag} algorithm \citep{Hancock12}\footnote{\url{https://github.com/PaulHancock/Aegean}}. This allows for comparison between the two programs (with {\Ag} being newer and still in development) and evaluation of the strengths and weaknesses of each. There are many algorithms we could have chosen to use; with the scope of this paper and to simplify the analysis we restrict the number of different source-extraction codes to two. We chose {\Ob} as it has been tested previously with VLA data. \textsc{Fndsou}, being an updated version of the \textsc{VSAD} algorithm used by the NVSS survey \citep{Condon98}, has a long development history, and many options to use to customize the fit. {\Ag} was chosen because it is one that is being developed with for possible use with SKA Pathfinder surveys; already in use with the low frequency pathfinder the Murchison Widefield Array \citep[MWA, see e.g][]{Wayth15,Offringa16}. In \citet{Hancock12} {\Ag} was compared against several other source finders (\textsc{miriad}'s {imsad}, \textsc{sextractor}, \textsc{sfind}, and ASKAPsoft's \textsc{selavy}) and was found to perform better than the others for sources near the noise limit and for blended sources. These are two regimes that we are interested in investigating for sources and therefore decided to use {\Ag} for this analysis. 

To understand the fit uncertainties associated with our images, noise properties, and fitting software, we generated simulated images and ran them through our fitting procedure. All of the simulated images used in this paper were made in the image plane only; no simulated \textit{uv} data were generated. For real data there can certainly be \textit{uv}-data related issues that lead to artefacts generating non-Gaussian fluctuations in images. These can particularly affect completeness and false detections. The VLA data we are addressing here (and in Paper \rom{2}) have good {\it uv} coverage, synthesized beam sidelobe peaks below $1\,$per cent, and are quite free from artefacts \citep[figure 3 from][shows how the noise closely matches a Gaussian distribution]{Condon12}. 

The use of simulated {\it uv} data can generate large data sets. Furthermore, since a {\it uv} sample corresponds to a particular baseline, then {\it uv}-based simulations would not necessarily apply well to all telescope configurations, and might be hard to scale up, in terms of data volume and computational power required. Different telescopes can introduce different types of artefacts, which may require accurate and or complicated models to reproduce. Additionally, going to the {\it uv} plane then back to the image plane opens up all of the different options for imaging that can impact the catalogue results. 

Given the VLA data considered here, the smaller scale of this work, and our aim to keep it more generalizable we felt that the addition of \textit{uv}-simulated data is unnecessary for this current study, although it will be important to investigate when considering imaging and cataloguing for larger surveys \citep[see e.g.][for examples of simulation studies using {\it uv} data]{Hopkins15,Huynh12}. This study examines the more simple and manageable case of having images with purely Gaussian instrumental noise. 

With the results from fitting the simulated images we were able to compare the fit output with the known ``true" input. We performed several sets of simulations: 
\begin{itemize}
\item simple -- test for effect of correlated noise in fit parameters;
\item realistic  -- test software accuracies with realistic flux and source size distribution as well as clustering;
\item sizes -- test specifically the effect of source size on the fit results;
\item blending -- test the effect of source blending on the fit results.
\end{itemize}

First we ran the simplest case of inserting point sources, with set flux densities of 4, 5, 8, 12, 20, and $100 \, \mu$Jy. These were inserted at points chosen from a grid over the image, such that the distance between the sources was $\gg \theta_{\rm B}$. These point sources were convolved with the beams from our C and CB images, then added to backgrounds of random Gaussian beam-convolved noise with $\sigma=1.08$ and $1.15\, \mu$Jy beam$^{-1}$ for the two resolutions. No primary beam correction was applied for these simulations. This initial case was used simply to determine the uncertainties of fitting in the presence of correlated noise.\footnote{By correlated noise we mean that the noise in adjacent pixels is not statistically independent. This is due to the effective convolution with the synthesized beam during the imaging process.} as well as the effect of the imposed size constraint. We created eight simulated images (each made at both the better and worse resolutions), each image having 400 sources in it. These images were then fit using {\Ob} and \Ag. The outputs were cross-matched with the true positions using a $5\,$arcsec search radius. Then the data from all eight cases were collated to look at the results. 

The images (without primary beam corrections) were searched for peaks down to the $3\sigma$ level. Each peak area was then fit with a 2D elliptical Gaussian, with the fit parameters being \speak, the centre RA and Dec positions, the major axis size \thetamaj, minor axis size \thetamin, and the position angle \posang. The fitting with {\Ob} was constrained so that the major and minor axes could not be less than the image beam size \thetabeam,\footnote{When the term `image beam' is used in referring to these data it indicates the synthesized beam rather than the primary beam.} this option not being available with \Ag. The total flux density {\stot} is computed as 
\begin{equation}
S_{\rm total}=S_{\rm peak} \frac{\theta_{\rm maj} \theta_{\rm min}}{\theta_{\rm B}^2}.
\label{eq:cat_sint}
\end{equation}

However, the real world case is more complicated. We next  carried out tests with as realistic a setup as possible. We used the source catalogues from eight separate $0.5\, {\rm deg} \, \times 0.5\,$deg areas of the SKADS S$^3$ simulation \citep[][with the full simulation being $20\,$deg$^2$ at $1.4\,$GHz]{Wilman08}. We know already that the source count from these simulations matches fairly closely to published $1.4\,$GHz counts, and it includes clustering, as well as a reasonable source size distribution. 

We scaled the flux densities from $1.4\,$GHz to $3\,$GHz using a spectral index $\alpha=-0.7$, where $S_{\nu_2}=S_{\nu_1}(\nu_2/\nu_1)^{\alpha}$. We then cut out any sources with $S>5\,$mJy to more closely approximate the Lockman Hole field, where the brightest source at $3\,$GHz within the primary beam has a peak flux density of $3\,$mJy beam$^{-1}$. We also cut out AGN objects with extended lobes or jets, because in real images the few such sources would not be fit with Gaussians (but handled on an individual basis) and we are solely concerned here with single-component sources. After the images were generated they were multiplied by the $3\,$GHz VLA primary beam power pattern. They were convolved with the CB and C beams to create better and worse resolution versions of each, and beam-convolved noise was added to each image. Small regions of one simulated image, at each resolution, are shown in Fig.~\ref{fig:simims}, and compared with regions of the same size from the real VLA images. These 16 images were then fit using {\Ob} and \Ag, with the outputs cross-matched with the known positions within a $5\,$arcsec search radius. For the remainder of the discussion, $S_{\rm peak}$ and $S_{\rm tot}$ refer to flux densities {\it without} a primary beam correction, unless otherwise noted.

\section{Parameter Uncertainties and Correction}
\label{sec:cat_simerr}

\subsection{Simple simulations}
\label{sec:simplesim}

In \citet{Condon97} the errors for 2D elliptical Gaussian fits are discussed. The square of the overall signal-to-noise ratio $\rho^2$ \citep[equation 20 from][]{Condon97} is
\begin{equation}
\rho^2=\frac{\pi}{8 \ln{2}} \frac{ \theta_{\rm maj} \theta_{\rm min} S_{\rm peak}^2}{ h^2 \mu^2},
\label{eq:cat_jimer1}
\end{equation}
where $h$ is the pixel size, and $\mu$ is the rms noise in each pixel of an un-smoothed image. The units of $\mu$ and $S_{\rm peak}$ here are Jy beam$^{-1}$ and the units of $h$ are arcsec$^2$.  The errors on each of the fit parameters, as related to $\rho^2$ \citep[equation 21 from][]{Condon97}, are
\begin{equation}
\begin{split}
 \frac{2}{\rho^2} & \simeq \frac{\sigma^2(S_{\rm peak})}{S_{\rm peak}^2}=\frac{\sigma^2(S_{\rm total})}{S_{\rm total}^2} \\ &= 8 \ln{2} \frac{\sigma^2(x_0)}{\theta_{\rm maj}^2} =8 \ln{2} \frac{\sigma^2(y_0)}{\theta_{\rm min}^2}\\& = \frac{\sigma^2(\theta_{\rm maj})}{\theta_{\rm maj}^2}  =\frac{\sigma^2(\theta_{\rm min})}{\theta_{\rm min}^2}  \\  &=\frac{\sigma^2(\phi)}{2} \left ( \frac{\theta_{\rm maj}^2-\theta_{\rm min}^2}{\theta_{\rm maj} \theta_{\rm min}}\right ) .
\end{split}
\label{eq:cat_jimer2}
\end{equation}
Here $x_0$, and $y_0$ are the source position coordinates and $\phi$ is the source position angle. 
\begin{figure*}
\includegraphics[scale=.535,natwidth=12in,natheight=12in]{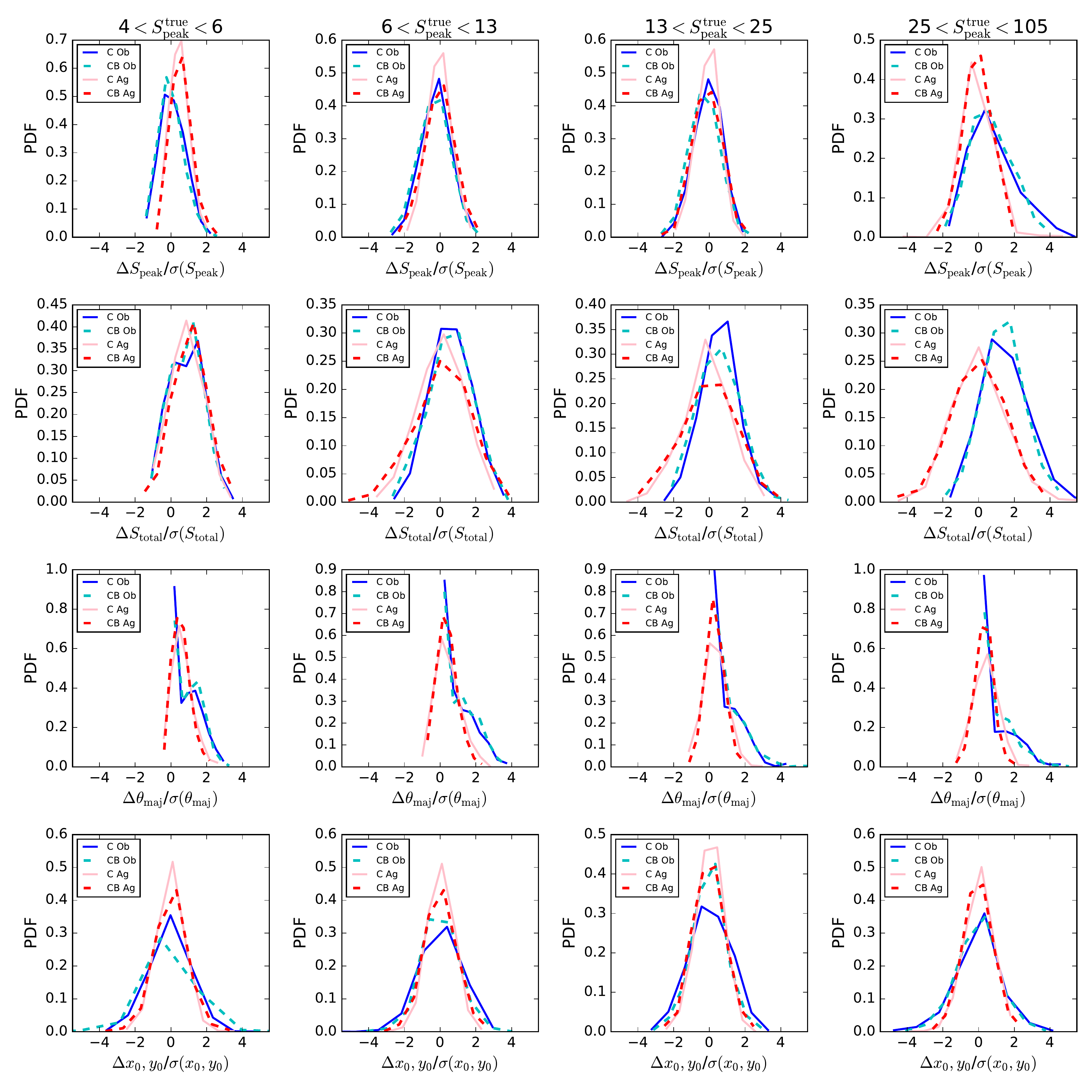}
\caption{Probability distributions of fitted parameter differences for the simple simulation. The parameter differences are computed in bins of true peak flux density, with the columns all being the same flux density bin and left to right are $4\le S \, [\mu {\rm Jy}] \le 6$, $6\le S \, [\mu {\rm Jy}] \le 13$, $13\le S \, [\mu {\rm Jy}] \le 25$, and $25\le S \, [\mu {\rm Jy}] \le 105$. All the values are normalized by the predicted uncertainties from equation~(\ref{eq:cat_jimer3}). The blue lines are fit with {\Ob}, dark blue solid is for C images and light blue dashed is for the CB. The light red solid lines are for the C images fit with {\Ag} and the dark red dashed lines are for the CB images fit with \Ag. From top to bottom the rows are $\Delta S_{\rm peak}/\sigma(S_{\rm peak})$, $\Delta S_{\rm total}/\sigma(S_{\rm total})$, $\Delta \theta_{\rm maj}/\sigma(\theta_{\rm maj})$, and $\Delta x_0,y_0/\sigma(x_0,y_0)$. } 
\label{fig:cat_simpleh}
\end{figure*}

\begin{figure}
\includegraphics[scale=.48,natwidth=7in,natheight=13.5in]{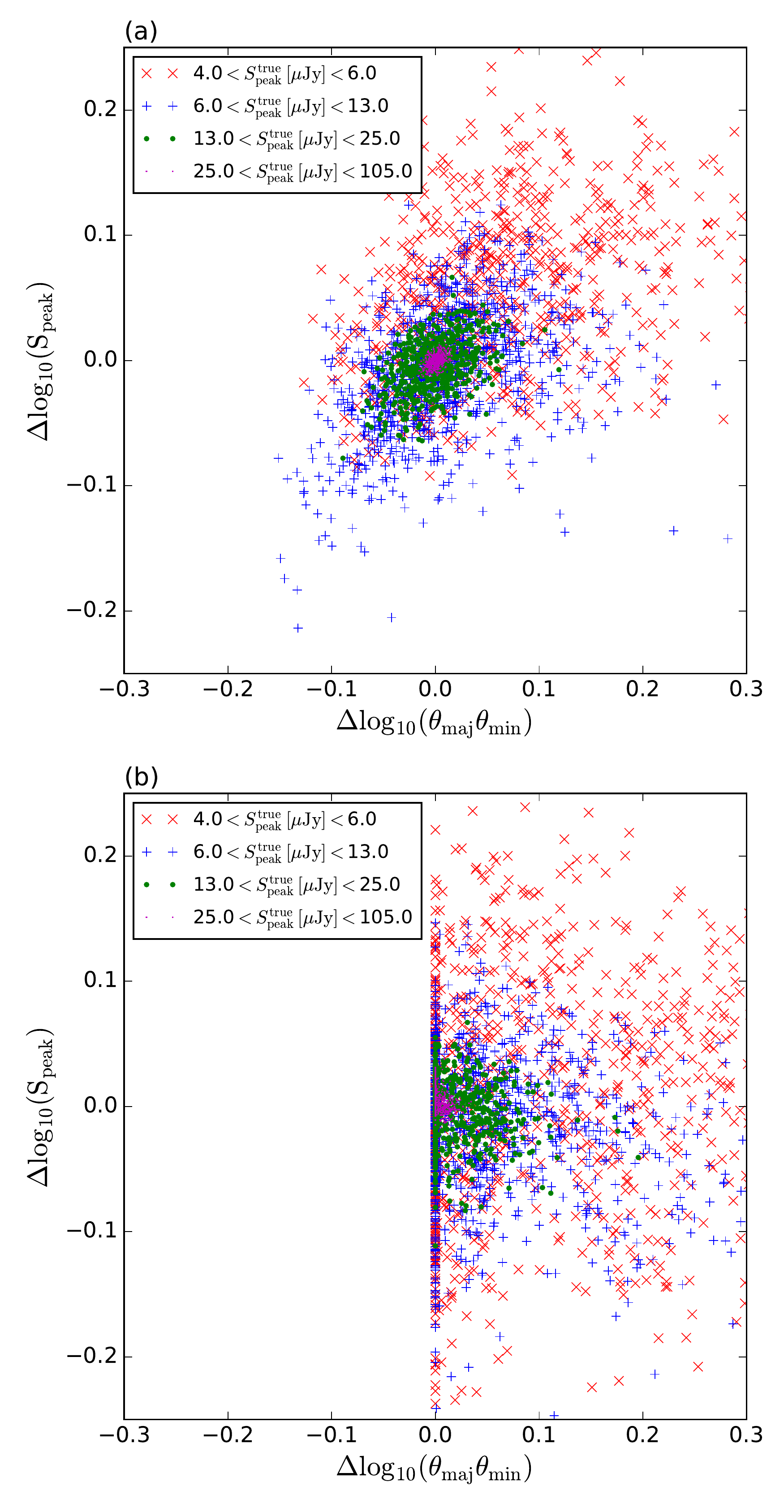}
\caption{Logarithmic difference in peaks compared to the logarithmic difference in fitted sizes for different peak flux densities. Panel (a) is for the C images with \Ag, while panel (b) is for the C images with \Ob.} 
\label{fig:cat_simplecr}
\end{figure}

The simple simulation case should test whether these equations hold up when the noise has also been convolved with the beam, and, in the {\Ob} case, the effect when the size constraint of being larger than the beam is imposed. According to \citet{Condon97}, for correlated noise, equation~(\ref{eq:cat_jimer1}) changes to
\begin{equation}
\scriptsize
\rho^2=\frac{ \theta_{\rm maj} \theta_{\rm min} S_{\rm peak}^2}{4 \theta_{\rm B}^2 \mu^2}\left [ 1+ \left( \frac{\theta_{\rm B}}{\theta_{\rm maj}} \right )^2 \right ]^{\beta_{\rm maj}}\left [ 1+ \left( \frac{\theta_{\rm B}}{\theta_{\rm min}} \right )^2 \right ]^{\beta_{\rm min}},
\label{eq:cat_jimer3}
\end{equation}
where $\mu$ is now the rms noise in a smoothed image and $\beta_{\rm maj}$ and $\beta_{\rm min}$ are to be determined by simulation.

Taking the fitted and matched data and binning by true flux density we computed the difference of fit to true peak flux density, total flux density, positions, major axis sizes, as well as the predicted uncertainty for each source from equation~(\ref{eq:cat_jimer3}). Histograms of these differences divided by the predicted uncertainties are shown in Fig.~\ref{fig:cat_simpleh}. This normalization is such that if the distributions perfectly matched the predictions the distributions should resemble a unit Gaussian distribution \citep{Condon97}. 

The distributions should be approximately unit Gaussians if equation~(\ref{eq:cat_jimer3}) holds, depending on the values used for $\beta_{\rm maj}$ and $\beta_{\rm min}$. We fit each distribution with a 1D Gaussian model and it appears that equation~(\ref{eq:cat_jimer3}) works well if $\beta_{\rm maj}=\beta_{\rm min}=1.25$ for the axis sizes and $\beta_{\rm maj}=\beta_{\rm min}=1.0$ for the flux densities and positional uncertainties. The mean of the standard deviations of the fitted 1D Gaussian models is $\simeq 1\pm.0.2$. 

The lowest flux density bins (the first column of Fig.~{\ref{fig:cat_simpleh}) show the largest positive shift in the means of the distributions. The peak flux densities and major axis sizes are being overestimated, and consequently the total flux densities. This is a known complication with most fitting routines when dealing with low signal-to-noise sources. The peak overestimation is less at higher flux densities while the sizes are still shifted, implying that the size discrepancy is the larger contributor to the total flux density overestimation.

The positional uncertainty for {\Ob} is slightly larger than that of \Ag, particularly for the worse resolution C images. The major axis distributions are one-sided for the {\Ob} fits, as expected since the constraint of a minimum size of the beam was imposed. The areas around the peaks match the {\Ag} fits closely; however, in the 1 to 4$\sigma$ region the {\Ob} distributions show an increase or bump across all flux densities. This is likely the reason why the $S_{\rm total}$ distributions for {\Ob} appear to be shifted from zero mean to a positive mean. This means that imposing the constraint of a minimum source size of the beam creates a bias of overestimated total flux-density values.

According to \citet{Condon97} if the image noise is uncorrelated, the fitted peak should be anti-correlated with the fitted angular sizes $\theta_{\rm maj}$ and $\theta_{\rm min}$, as well as their product $\theta_{\rm maj}\theta_{\rm min}$. Figure~\ref{fig:cat_simplecr} shows the change in peak flux densities against the change in the size product for the C image resolution for different flux density ranges. Panel (a) of Fig.~\ref{fig:cat_simplecr}, fit results from {\Ag}, shows these values to be positively correlated, rather than anti-correlated. Panel (b), for {\Ob}, is only one-sided due to the size constraint in the fitting, but does not seem to show any positive or negative correlation. The presence of correlated noise appears to create a positive correlation between the fitting of the peak flux density and axis sizes, which negated, or lessened, by introducing the size constraint. 

\subsection{Realistic simulations}
\label{sec:realsim}

\begin{figure*}
\includegraphics[scale=.535,natwidth=12in,natheight=2in]{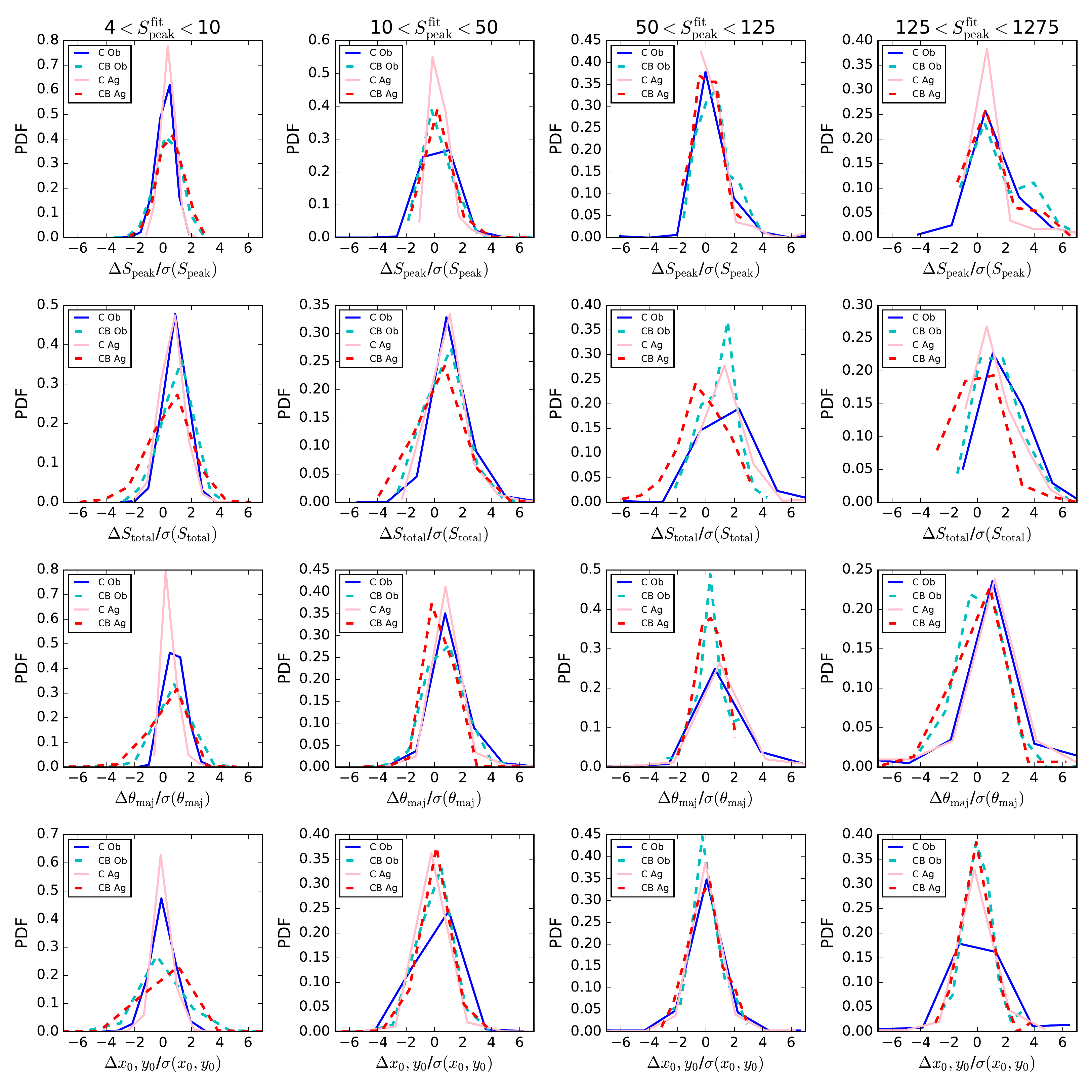}
\caption{Probability distributions of fitted parameter differences for the realistic simulation. The parameter differences are computed in bins of fitted peak flux density, with the columns all being the same flux density bin and left to right are $4\le S_{\rm peak}^{\rm fit} \, [\mu {\rm Jy}] \le 10$, $10\le S_{\rm peak}^{\rm fit} \, [\mu {\rm Jy}] \le 50$, $50\le S_{\rm peak}^{\rm fit} \, [\mu {\rm Jy}] \le 125$, and $125\le S_{\rm peak}^{\rm fit} \, [\mu {\rm Jy}] \le 1275$. All the values are normalized by the predicted uncertainties from equation~(\ref{eq:cat_jimer3}). The blue lines are fit with {\Ob}, dark blue solid is for C images and light blue dashed is for the CB. The light red solid lines are for the C images fit with {\Ag} and the dark red dashed lines are for the CB images fit with \Ag. From top to bottom the rows are $\Delta S_{\rm peak}/\sigma(S_{\rm peak})$, $\Delta S_{\rm total}/\sigma(S_{\rm total})$, $\Delta \theta_{\rm maj}/\sigma(\theta_{\rm maj})$, and $\Delta x_0,y_0/\sigma(x_0,y_0)$.}
\label{fig:cat_simerrsh}
\end{figure*}

\begin{figure}
\includegraphics[scale=.33,natwidth=10.75in,natheight=9in]{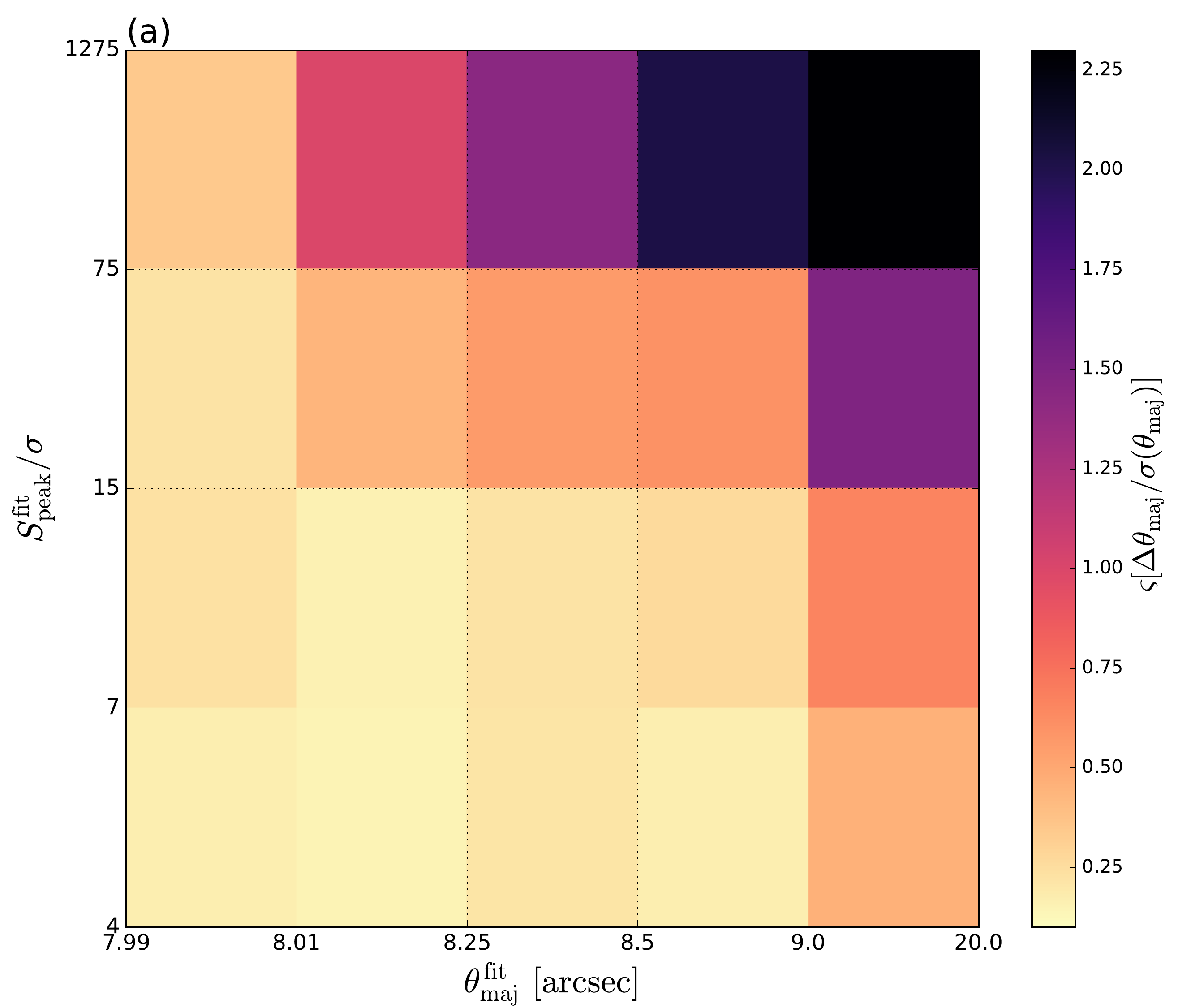}
\includegraphics[scale=.33,natwidth=10.75in,natheight=9in]{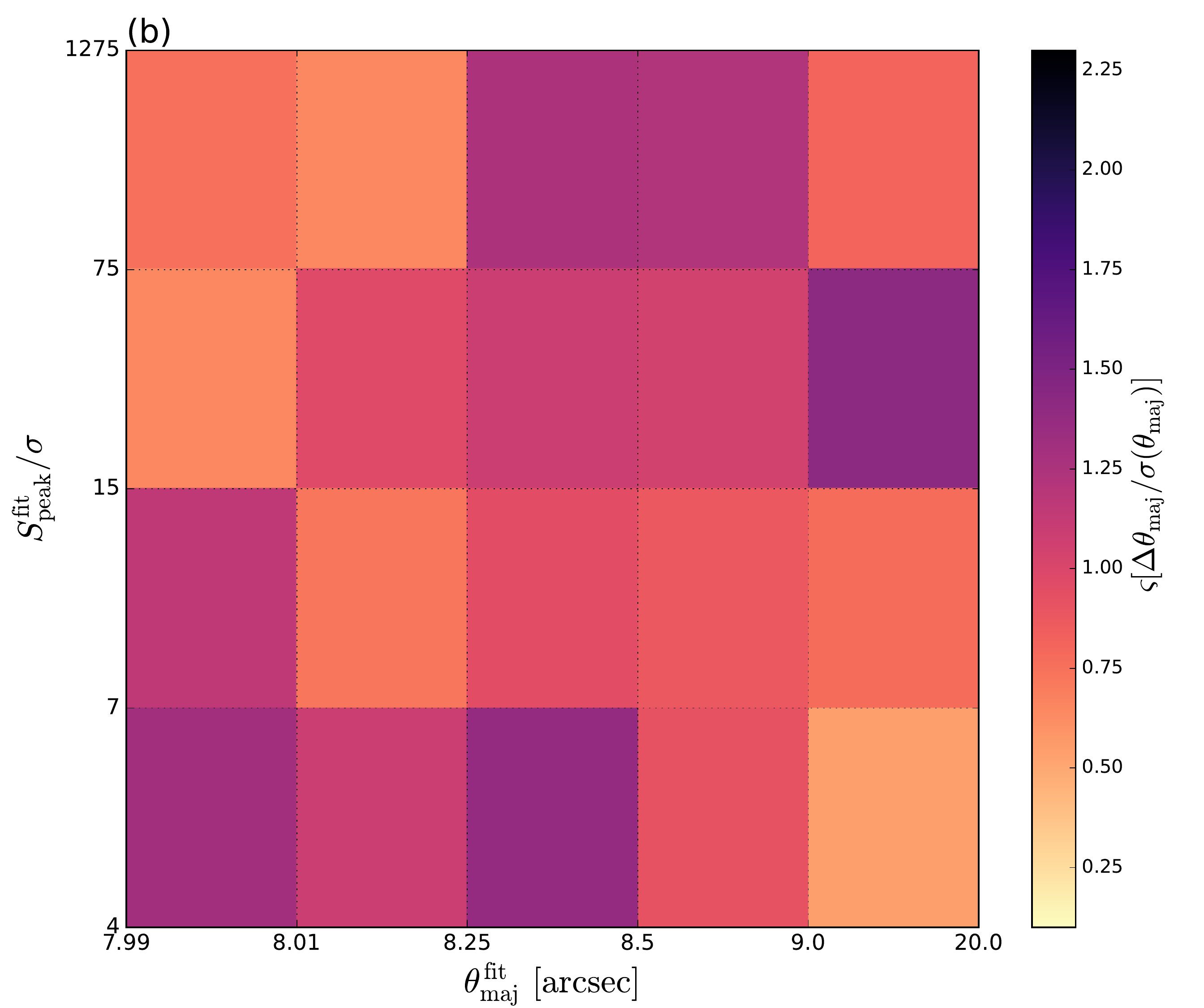}
\caption{Two dimensional distribution of standard deviations $\varsigma$ of $\Delta \theta_{\rm maj}/\sigma(\theta_{\rm maj})$ for C-resolution images fit with the {\Ob} software in bins of peak fitted flux density and fitted major axis. The leftmost bins are unresolved sources ($\theta_{\rm maj}=\theta_{\rm B}$). Panel (a) shows the distribution computed using equation~(\ref{eq:cat_jimer3}) with values of $\beta_{\rm maj}=\beta_{\rm min}=1.25$. Panel (b) shows the distribution computed using equation~(\ref{eq:cat_jimer3}), but using equations~(\ref{eq:alphamj}) and (\ref{eq:alphamn}) for the $\alpha$ values, with $k_{\rm maj}=7.47$ and $k_{\rm min}=2.89$.}
\label{fig:majerr}
\end{figure}

The realistic simulation differs from the simple case in that we now add in the issues of the primary beam, confusion noise, and clustering or source blending, as well as a range of source flux densities and sizes. The lower-resolution images should be much more affected by confusion and clustering.

We took all the sources from the realistic simulations that were found from the fitting routines and cross-matched that catalogue with the true, or input, source catalogue. Using the matches, we computed the differences of fit to true and the ratios of fit to true for all the fit parameters. These ratios and differences were binned by their fitted peak flux densities. Figure~\ref{fig:cat_simerrsh} shows the probability distributions normalized by the predicted uncertainties of equation~(\ref{eq:cat_jimer3}) for the flux density bins, for both image resolutions, using the values of $\beta_{\rm maj}$ and $\beta_{\rm min}$ obtained in Sec.~\ref{sec:simplesim}. For $\mu$ in equation~(\ref{eq:cat_jimer3}) instead of the 1.1 and $1.15\, \mu$Jy beam$^{-1}$ for the C and CB images, we use the values of 1.6 and $1.35\, \mu$Jy beam$^{-1}$. These values were obtained by adding the instrumental noise and confusion noise in quadrature. 

We can see the better-resolution distributions in Fig.~\ref{fig:cat_simerrsh} look similar to those in Fig.~\ref{fig:cat_simpleh}, whereas the low resolution distributions have been widened and, in some cases, shifted. Fitting 1D Gaussian models yields standard deviations of approximately $1.0\pm0.3$. For \speak, \stot, and {\thetamaj} the peak of the distributions either have a positive shift in the mean, a larger positive tail, or both (for the CB distributions as well in some cases). This implies the fitting tends to overestimate the parameters, as was seen with the simple simulation case, although in this case the sources are also affected by confusion or blending, not present in the simple simulations. 

Equation~(\ref{eq:cat_jimer3}) with set values of $\beta_{\rm maj}$ and $\beta_{\rm min}$ works well in most cases. However, if we bin the data by peak flux densities and major-axis fitted sizes and compute the standard deviations of $\Delta \theta_{\rm maj}/\sigma(\theta_{\rm maj})$, we can see how the accuracy of $\sigma(\theta_{\rm maj})$ varies by brightness and size. It turns out the uncertainties are overestimated for faint flux densities and small axis sizes and underestimated for brighter sources with larger sizes. The distribution of standard deviations $\varsigma$ of $\Delta \theta_{\rm maj}/\sigma(\theta_{\rm maj})$ is shown in panel (a) of Fig.~\ref{fig:majerr} for the C image {\Ob} results, where the mean standard deviation is $\langle \varsigma \rangle=0.64\pm0.6$. It seems that $\beta_{\rm maj}$ and $\beta_{\rm min}$ must vary according to the peak flux densities and sizes. Rather than just fitting for different $\beta$ values at set intervals, we derived $\beta$'s as functions of $S_{\rm peak}$, $\theta_{\rm maj}$, and $\theta_{\rm min}$. We find that
\begin{equation}
\beta_{\rm maj}=k_{\rm maj}-\left ( \frac{\theta_{\rm maj}}{\theta_{\rm B}} \right ) ^2 \log_{10}\left [ \left( \frac{S_{\rm peak}}{\sigma} \right ) ^2 \right],
\label{eq:alphamj}
\end{equation}
and
\begin{equation}
\beta_{\rm min}=k_{\rm min}-\left ( \frac{\theta_{\rm min}}{\theta_{\rm B}} \right ) ^2 \log_{10}\left [ \left( \frac{S_{\rm peak}}{\sigma} \right ) ^2 \right],
\label{eq:alphamn}
\end{equation}
can correct for this effect. In these equations $k_{\rm maj}$ and $k_{\rm min}$ are constants to be determined for a given data set. For the C-image results of {\Ob} we used least-squares fitting to find the values which bring all the standard deviations closest to one, finding $k_{\rm maj}=7.47$ and $k_{\rm min}=2.89$ for {\Ob} C-images. This distribution is shown in panel (b) of Fig.~\ref{fig:majerr}, which has a mean of $\langle \varsigma \rangle =0.98\pm0.25$. We adopt these equations for the axis size uncertainties; however, for the other parameters, such as $S_{\rm peak}$, the set values of $\beta$ are still acceptable. 

The effect of the added size constraint with the {\Ob} output is harder to detect with these results, given that not all of the sources were unresolved. In this case, for the most part, the {\Ob} distributions are similar to the {\Ag} distributions, which have no size constraint. There may still be a bias, but with the added effects from resolved sources, confusion, and clustering it is not significant.

 \begin{figure}
\includegraphics[scale=.34,natwidth=10.75in,natheight=9in]{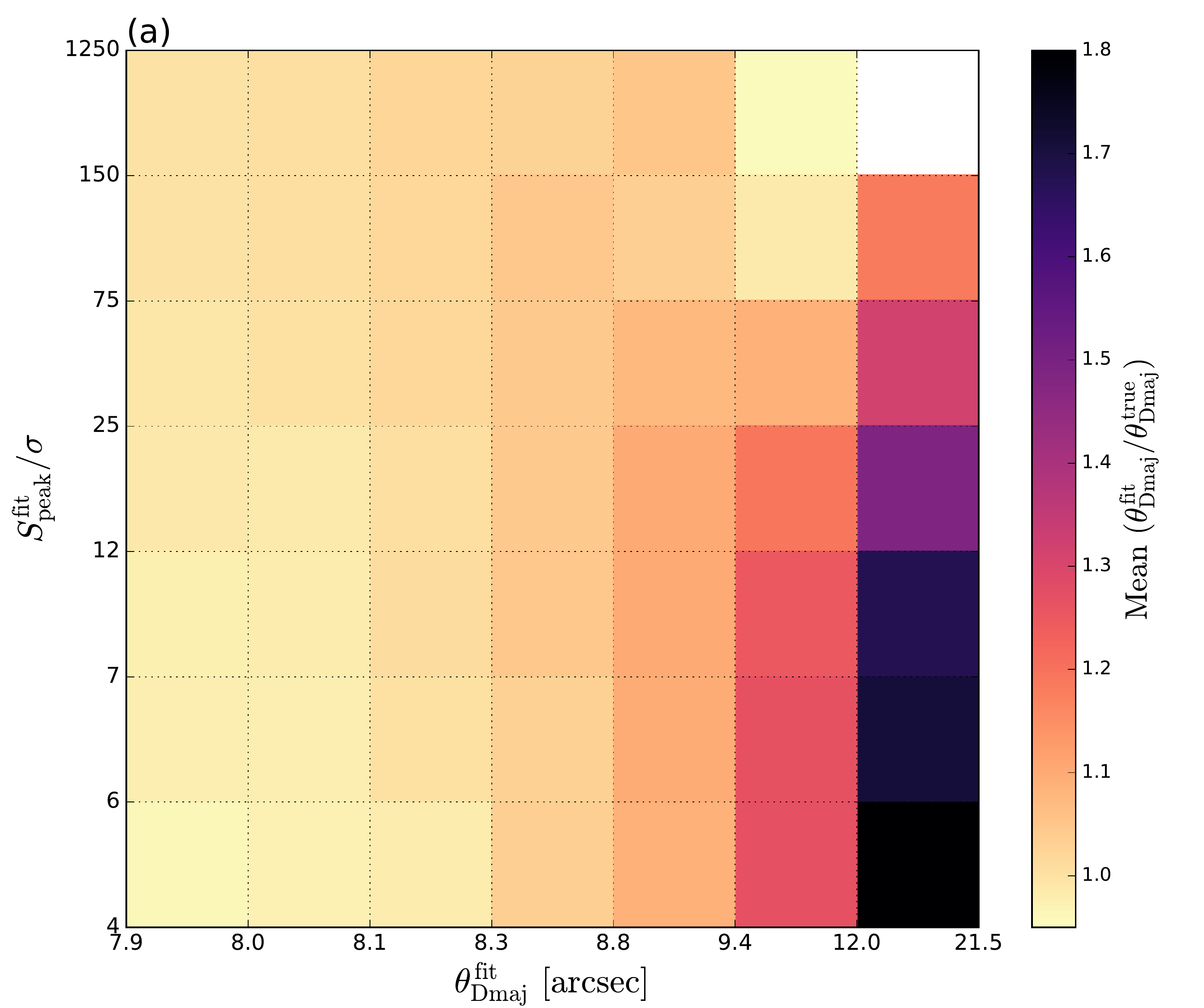}
\caption{Results of comparison of fitted to true values of source parameters for C-resolution images fit with the {\Ob} software. The shading shows the mean ratio of fitted deconvolved major axis size to true deconvolved major axis size of sources from the realistic simulation, computed in bins of peak fitted SNR and fitted major axis FWHM. The white bins indicate that there are no sources within those size and peak ranges. The leftmost bins are unresolved sources ($\theta_{\rm maj}=\theta_{\rm B}$).}
\label{fig:cat_2dinterp}
\end{figure}

\begin{figure*}
\includegraphics[scale=.55,natwidth=10in,natheight=10in]{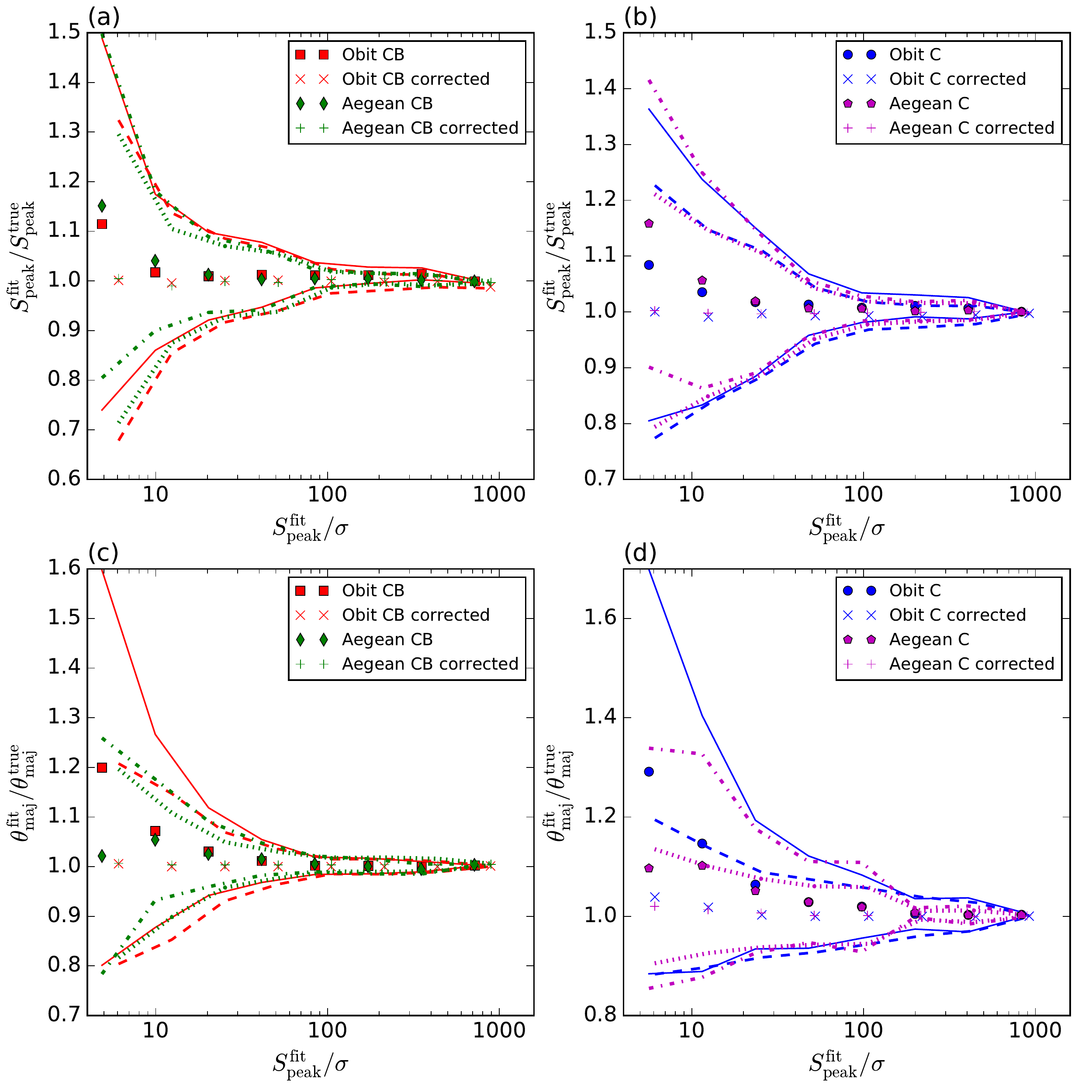}
\caption{Mean of the ratio of the fitted to true values for peaks (top row) and major axis size (bottom row), from the realistic simulations plotted in bins of fitted peak signal-to-noise ratio The ``CB" values were made with CB data image resolution (left column) and ``C" values made with C data image resolution (right column). The corrected values are after using the method of 2D binning and interpolation to obtain correction factors and uncertainties for individual sources. The points are the bin means, while the lines show the $\pm1\sigma$ uncertainty regions. The blue points and lines are C data fit with \Ob, while the red points and lines are CB data fit with \Ob, solid lines for no corrections and dashed lines for corrected values. The green points and lines are the CB data fit with {\Ag} and the purple points and lines are the C data fit with \Ag, with the dot-dashed lines uncorrected data and dotted lines corrected data.}
\label{fig:cat_simerrs2}
\end{figure*}

The goal of the more realistic simulations was to use the results in order to estimate corrections that can be applied to the real data. To do this we took all the matched simulated sources and binned them by fitted peak flux density and fitted deconvolved major axis size, where the deconvolved major axis {\thetadmaj} and deconvolved minor axis {\thetadmin} are computed as $\theta_{\rm Dmaj}=\sqrt{\theta_{\rm maj}^2-\theta_{\rm B}^2}$ and $\theta_{\rm Dmin}=\sqrt{\theta_{\rm min}^2-\theta_{\rm B}^2}$. We computed the mean and standard deviations of the ratios of the fitted values to the true values, i.e. $S_{\rm peak}^{\rm fit}/S_{\rm peak}^{\rm true}$ in two dimensional bins. This was done rather than just binning by peak flux density, since those sources with large fitted axes tend to have larger offsets and uncertainties compared with those with sizes closer to the beam size. An example for the mean ratios of the major axis for the C image {\Ob} results is shown in Fig.~\ref{fig:cat_2dinterp}.

The addition of binning by size shows that the effect of overestimation in a parameter is largest for those sources not only having the faintest peaks, but also with the largest sizes. Figure~\ref{fig:cat_2dinterp} specifically shows the 2D array for correction of the major axis size; similar results were also found for the fitted peaks and minor axis sizes. The bin setups were chosen to ensure a minimum of $10$ sources in a bin (or genuinely zero in some cases). 

The arrays shown in Fig.~\ref{fig:cat_2dinterp} can be used to compute individual source-parameter corrections. By interpolating the value for each source from the mean ratio arrays, one can make an adjustment for each parameter. We tested this correction method for all of the simulated sources, interpolating corrections for peak flux density, total flux density, major and minor axis deconvolved sizes. Figure~\ref{fig:cat_simerrs2} shows how the mean ratios and uncertainties change after interpolating new values for the fit parameters. We can see that for the faintest flux density sources (where previously the mean ratios were greater than one) after correction, are all now much closer to unity.  

This method enables us to account for the effect known as ``flux boosting" \citep{Jauncey68,Coppin05} in our catalogue. This well known effect means that, particularly for sources near the noise limit, the fitted flux densities are more likely to be overestimated than underestimated, due to both instrumental and confusion noise, this results in an overestimate of the number of bright sources. There are many papers that discuss correcting for flux boosting using Bayesian methods \citep[e.g.][]{Jauncey68,Coppin05,Chapin11}. However, many of those studies deal with submm or infrared data, where $\theta_{\rm B}\gg \theta_{\rm s}$ ($\theta_{\rm s}$ being the source size) and therefore are only fitting for a total flux density rather than a peak together with size and shape. Submm and infrared images are also not affected by the primary beam, which complicates the Bayesian method used to compute probabilities directly from source counts for interferometric data. Therefore, rather than adopting a Bayesian deboosting prescription, we chose to use the realistic simulations and this 2D interpolation to account for these effects by computing the corrections for both size axes and for the peak flux densities.

\section{Completeness and False Detections}
\label{sec:cat_cf}
The simulated catalogues can also be used to examine how well the fitting software does at finding all of the known sources, as well as how often it adds false detections.
\begin{figure*}
\includegraphics[width=6.89in,height=5.05in,natwidth=13in,natheight=9.5in]{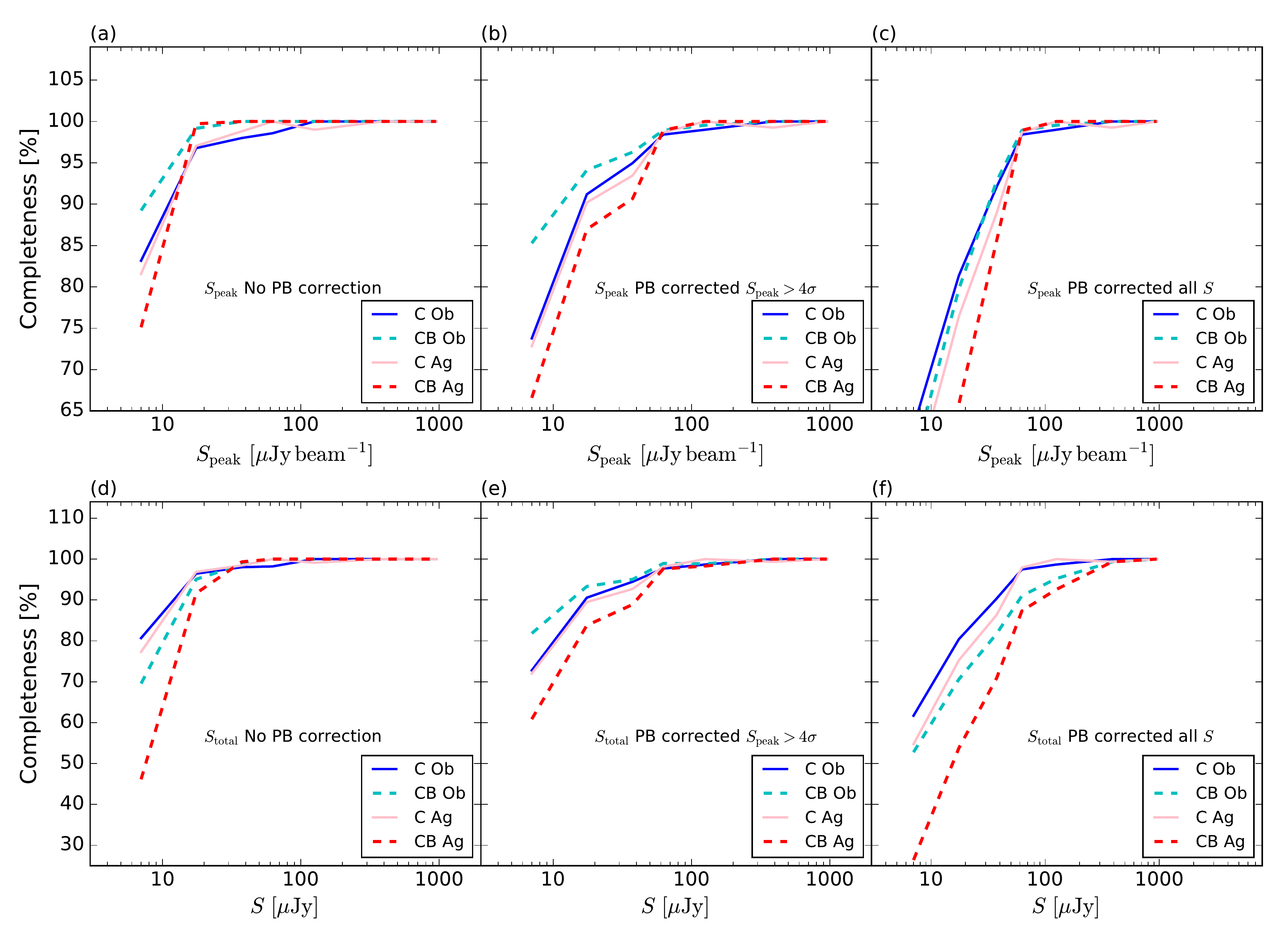}
\caption{Completeness of sources derived from simulations. The plots show the number of sources binned by flux density (of the sources with matches) divided by the true total number of inserted sources. The top row shows the sources binned by peak flux density, while the bottom row shows them binned by total flux density. Panels (a) and (d) show the sources binned by flux density uncorrected for the primary beam. Panels (b) and (e) are for primary-beam-corrected flux densities, but only including those sources whose uncorrected peaks are $>4\sigma$, i.e. those that could have been detected. Panels (c) and (f) are the primary-beam-corrected flux density including all sources. The solid dark blue line is the C image results from \Ob, the dashed light blue dashed line is the CB image results from \Ob, the light red solid line is the C image results from \Ag, and the dark red dashed line is the results from the CB image from \Ag. }
\label{fig:cat_simcomplete}
\end{figure*}

\begin{figure*}
\includegraphics[width=6.89in,height=5.05in,natwidth=13in,natheight=9.5in]{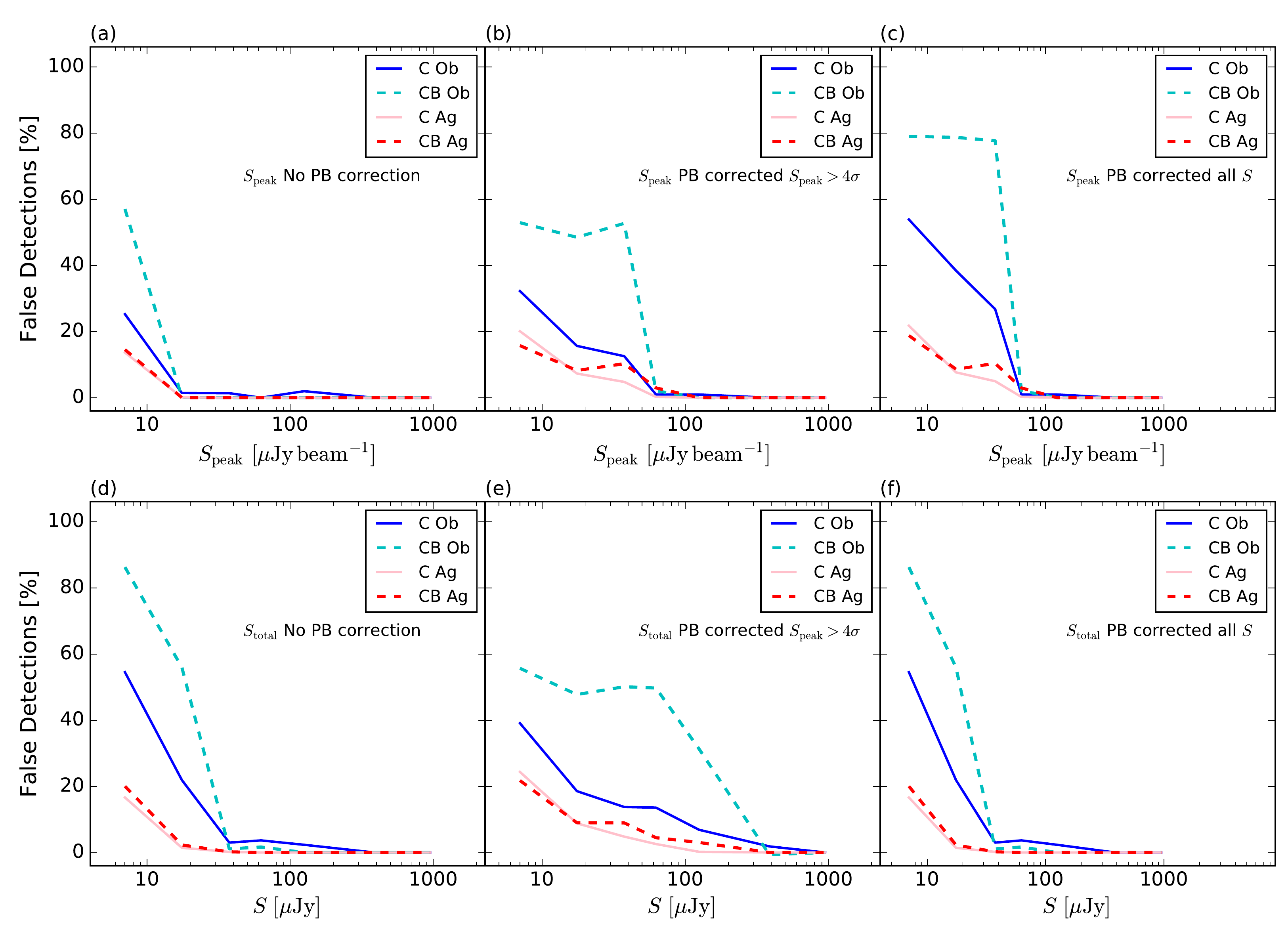}
\caption{Percentage of false detections of sources from simulations. The plots show one minus the ratio of matches to all fit sources binned by flux density. The top row shows the sources binned by peak flux density, while the bottom row shows them binned by total flux density. Panels (a) and (d) show the sources binned by flux density uncorrected for the primary beam. Panels (b) and (e) are for primary-beam-corrected flux densities, but only including those sources whose uncorrected peaks are $>4\sigma$, i.e. those that could have been detected. Panels (c) and (f) show the primary-beam-corrected flux density including all sources. The solid dark blue line shows the C image results from \Ob, the dashed light blue dashed line shows the CB image results from \Ob, the light red solid line shows the C image results from \Ag, and the dark red dashed line shows the results from the CB image from \Ag.}
\label{fig:cat_simcfalse}
\end{figure*}

The issues of completeness and false detections are somewhat complicated by whether one is interested in uncorrected or primary beam-corrected data, peak or total flux densities, and whether we look at the fitted values or the true values of the matched sources. Looking at the uncorrected peak values is best when analysing how well the fitting software performs; the primary beam corrected total flux density values should be used to obtain corrections to apply to a source count. 

Examples of this for completeness are shown in Fig.~\ref{fig:cat_simcomplete}. These plots show the number of sources, in bins of flux density, that were found by the fitting routine and that had a known match within $5\,$arcsec, divided by the total number of known inserted objects in that flux density bin. This shows that the completeness levels are lower if we consider the total flux density values rather than the peak flux densities. This difference is due to the source detection process looking at the peak pixel values. Accordingly, objects with high total flux density but extended sizes will have lower peak fluxes and tend more often to fall below the detection threshold. 

The figure also shows the difference that the primary beam and source size makes. For example, in the CB image for total flux densities at the faintest levels, when considering those that could have been detected (i.e. peaks $>4\sigma$, bottom middle panel of Fig.~\ref{fig:cat_simcomplete}) the completeness is roughly $85\,$per cent, whereas it drops closer to $65\,$per cent when considering all the known sources whose total flux density is above $4\sigma$ (including ones whose peak is below $4\sigma$, panel (f)). 

Examples of the false detection rate found in the simulations are shown in Fig.~\ref{fig:cat_simcfalse}. The plots show the total number of detected sources minus the number of sources with matches expressed as a percentage, binned by flux density. Again we see how the false detection rate drops quickly to zero when considering non-primary beam-corrected flux densities, but stays higher to higher flux densities when looking at the corrected flux densities, which is due to the uncorrected false sources being shifted into higher flux density bins by the primary beam correction). There are a lower number of false detections for the C images, due to its lower instrumental noise. 

It is clear from Fig.~\ref{fig:cat_simcfalse} that {\Ob} performs worse when it comes to false detections at both resolutions, with the better resolution case reaching almost $80\,$per cent false detection rate for the faint flux densities. {\Ob}'s \textsc{Fndsou} does have a feature for incorporating the estimated false detection rate (FDR). \footnote{For full details of the implementation see \url{ftp://ftp.cv.nrao.edu/NRAO-staff/bcotton/Obit/FDR.pdf}} {\Ob} allows the user to set the maximum fractional false detection rate, FDR$_x$. It uses the image pixel histogram to calculate
\begin{equation}
{\rm FDR}_x=1-\frac{n_+ - n_-}{n_+},
\label{eq:fdr}
\end{equation}
where $n_+$ is the number of positive pixels in the histogram of bin $x$ and $n_-$ is the number of pixels in the bin of $-x$. The user can also decide what sized region around a source (or the whole image) that the code should use when making the histogram. 

As a test we took one of the simulated images with the CB resolution and reran the {\Ob} fitting with different levels set for the maximum acceptable false detection rate. We used FDR values of 0.01, 0.025, 0.05, 0.075, 0.1, and 0.25 and compared the results with the case of no FDR constraint. Figure ~\ref{fig:fdr} shows the completeness and false detection percentages for the different cases. The results show improvement over the case where no constraint is used. {\Ob} warns against using values that are too small ($< 0.05$) as it requires a large number of pixels to obtain the needed accuracy and can reject real sources. This can be seen in Fig.~\ref{fig:fdr}, where the FDR$=0.01$ case results in a much lower false detection rate but also lower completeness levels than the case of no constraint. Setting FDR too high can also prove troublesome. The FDR$=0.25$ case resulted in higher false detection rates for faint flux densities than the no constraint case. The middle values of $0.05\le FDR \le 0.10$ seem to produce the best results. The false detection percentage is lowered by roughly $20\,$per cent, with the completeness level increased by $5\,$per cent (on average). The {\Ob} FDR feature does produce better results, although the parameters require some caution and fine tuning.  

These simulation results are used to apply corrections to the direct source count of the real observations in Paper \rom{2}.  

\begin{figure}
\includegraphics[scale=.48,natwidth=7in,natheight=13in]{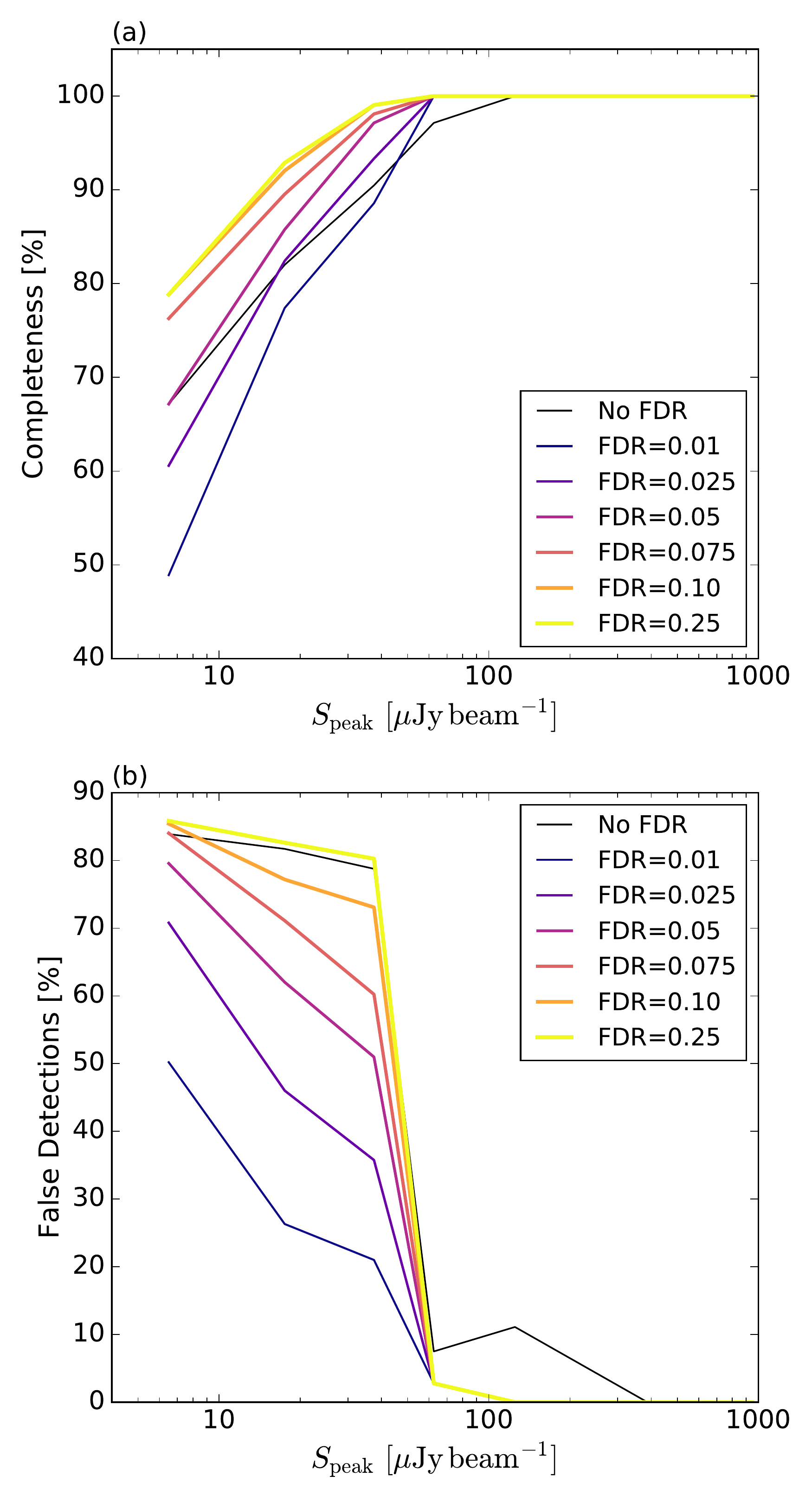}
\caption{Completeness and false detection results of fitting one simulated CB-resolution image with {\Ob}'s false detection rate constraint. Panel (a) shows the completeness percentage in bins of primary-beam-corrected peak flux density and panel (b) is the percentage of false detections. The maximum acceptable fractional false detection rate (FDR) values used in the fitting are: no constraint, 0.01, 0.025, 0.05, 0.075, 0.1, and 0.25 (indicated by the strength and colour of the lines). }
\label{fig:fdr}
\end{figure}

\section{Source Size}
\label{sec:sizes}

The issue of source angular size is particularly important to understand in interferometric data. A source's size has an effect on whether the source is detected and how well its parameters are fit.  A large source size compared to the beam for sources near the noise limit may have a beam-convolved peak too low to be detected by the fitting software.  For sources with large sizes compared to the beam that are detected, some emission may be resolved out, likely resulting in underestimated flux densities. 

To examine this effect in more detail we created a separate set of simulated images. We generated 500 sources with random positions and random total flux densities within the range $4\le S_{\rm total} \le 100 \, \mu$Jy.  The positions were chosen so that no source was closer than $40 \,$arcsec to another source, to avoid source blending contamination. The sources were set initially as point sources, then four additional images were made changing each source to a Gaussian shape with major and minor axis sizes of 1.25, 2.5, 5.25, and $10\,$ arcsec. All five of these images were then convolved with both the C-image beam and the CB-image beam for a total of 10 images. Finally, (beam-convolved) Gaussian noise was added to the images. 

Source finding and fitting were then performed on these images with both {\Ob} and \Ag, as previously described. The output was cross-matched with the known input of each image using a search radius of $5\,$arcsec. The fractions of sources found within $5\,$arcsec of a true source for the different algorithms and resolutions are shown in Fig.~\ref{fig:sz_frac}.

The C image, with worse resolution, is less affected by source size, as expected. Both resolutions show a decrease in the fraction of matched sources when the source size is roughly equal to or greater than the beam size. For the $2.75\,$arcsec beam size of the CB image, this can mean a significant number of sources being missed or having increased errors in fitting due to source size. Panels (b) and (c) of Fig.~\ref{fig:sz_frac} show the completeness as a function of total and peak input flux densities for the CB data for {\Ob} and panels (d) and (e) show the completeness for the {\Ag} data. These indicate that fainter sources are missed more often, with all of the missing sources having peak flux densities below $10\, \mu$Jy beam$^{-1}$.

In terms of the sources that are found by the software, we can see how accurately they were fit, as a function of input source size. Figure~\ref{fig:sz_rats} shows a comparison of fit values to true values for different parameters as a function of source size for different total input flux densities. Again, the worse resolution images show little effect until the source size becomes larger than the beam size. For the better resolution images, the peaks tend to be overestimated, by up to $200\,$per cent for sizes of $10\,$ arcsec and total flux density between 15 and $50\, \mu$Jy and almost $500\,$per cent for source sizes of $10\,$arcsec and total input flux density between 3 and $15\, \mu$Jy. The sizes are underestimated. For a source size of $10\,$arcsec and input total flux density between 15 and $50\, \mu$Jy the size is underestimated by $75\,$per cent and for flux densities between 3 and $15\, \mu$Jy this drops to $50\,$per cent.

However, since the total flux density is calculated from \speak (\thetamaj \thetamin / \thetabeam$^2)$, if both {\thetamaj} and {\thetamin} are underestimated, this over-compensates for the overestimated peak and results in a large underestimation of the total flux density for sources with sizes greater than the beam; the total flux density for source sizes of $10\,$arcsec is underestimated by about $65\,$per cent for sources with total input flux densities $\le 50\, \mu$Jy. The accuracy of the peak position also declines in relation to source size. The positional uncertainty goes from near zero at source size zero to 2 to $3\,$arcsec, depending on flux density. The rise in fit errors as the source true size increases is quite dramatic. The source's true size lowers the peak (from what it would be if it were a point source) by a factor of ((\thetadmaj \thetadmin)/\thetabeam$^2$+1). For a source with true axis sizes of $10\,$arcsec and a $2.75\,$arcsec circular beam, this corresponds to a factor of roughly 15. Thus only sources with \stot$\ge 90\, \mu$Jy would be above a $5\sigma$ detection limit, without accounting for primary beam effects.

This simulation tells us that source size can have a large impact on accurately cataloguing sources, and in obtaining accurate source counts. Once the source size is larger than the image beam, $30$ to $70\,$per cent of the sources are not found by the fitting software. Those that are found are likely to have errors in the fit parameters, such that the estimated total flux density is less than it should be. Thus the number of sources as a function of total flux density is affected by missing sources, and by misfitting. This issue, while having a large effect on the source count, will also affect the determination of other characteristics, such as spectral indices (because of the wrong flux density at this frequency), cross-matching with other wavelengths (inaccurate source position), and estimates of the source size distribution.

For the case here of $\theta_{\rm B}=2.75\,$arcsec, the effects are more moderate. However, for images with $\theta_{\rm B}\le 2\,$arcsec this represents a larger problem. If data are only available at a single high resolution there is no good way of discriminating between an accurately fit source and one misfit due to a larger size. 

The only viable solution is to obtain comparably deep data at multiple resolutions, which enables one to compare fits of the same source at different resolutions and to determine the most accurate fit. This may be not be an issue for some telescope arrays with more complete {\it uv}-coverage. More evenly distributed \textit{uv}-coverage over a range of angular scales would allow for imaging of data at multiple resolutions, or with multi-scale imaging techniques, without much loss in sensitivity.

\begin{figure*}
\includegraphics[scale=.515,natwidth=13.8in,natheight=8.5in]{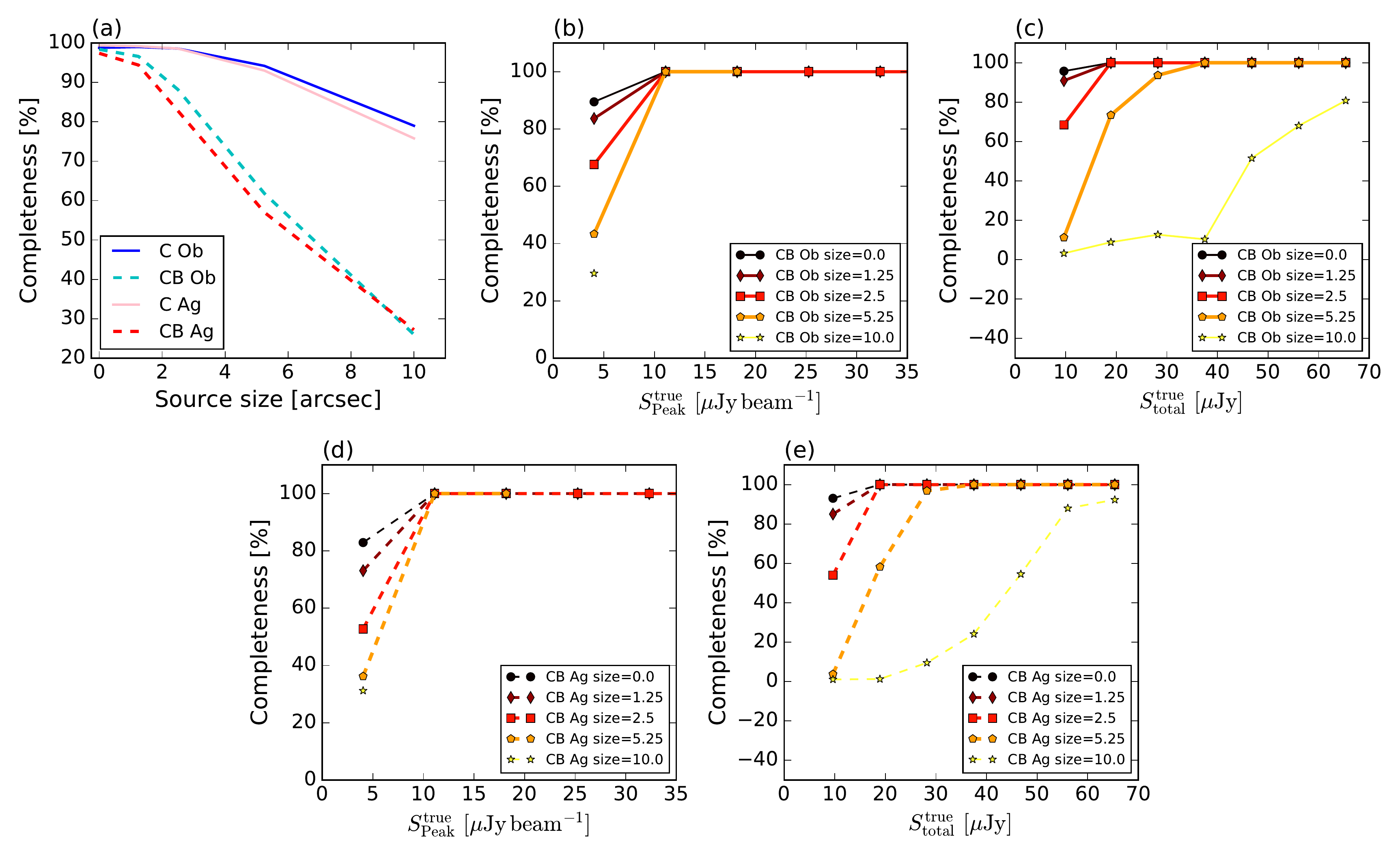}
\caption{Fraction of sources matched for different source sizes. The solid dark blue line is the C image results from \Ob, the dashed light blue dashed line is the CB image results from \Ob, the light red solid line is the C image results from \Ag, and the dark red dashed line is the results from the CB image from \Ag. Panel (a) shows the fraction as a function of true deconvolved source size.  Panel (b) is the fraction matched as a function of the true source peak flux density, also for the CB {\Ob} data. Panel (c) is the fraction matched as a function of true total flux density, shown for the CB {\Ob} data.
Panel (d) is the fraction matched as a function of the true source peak flux density, also for the CB {\Ag} data. Panel (e) is the fraction matched as a function of true total flux density, shown for the CB {\Ag} data.}
\label{fig:sz_frac}
\end{figure*}

\begin{figure*}
\includegraphics[scale=.47,natwidth=14.5in,natheight=7.75in]{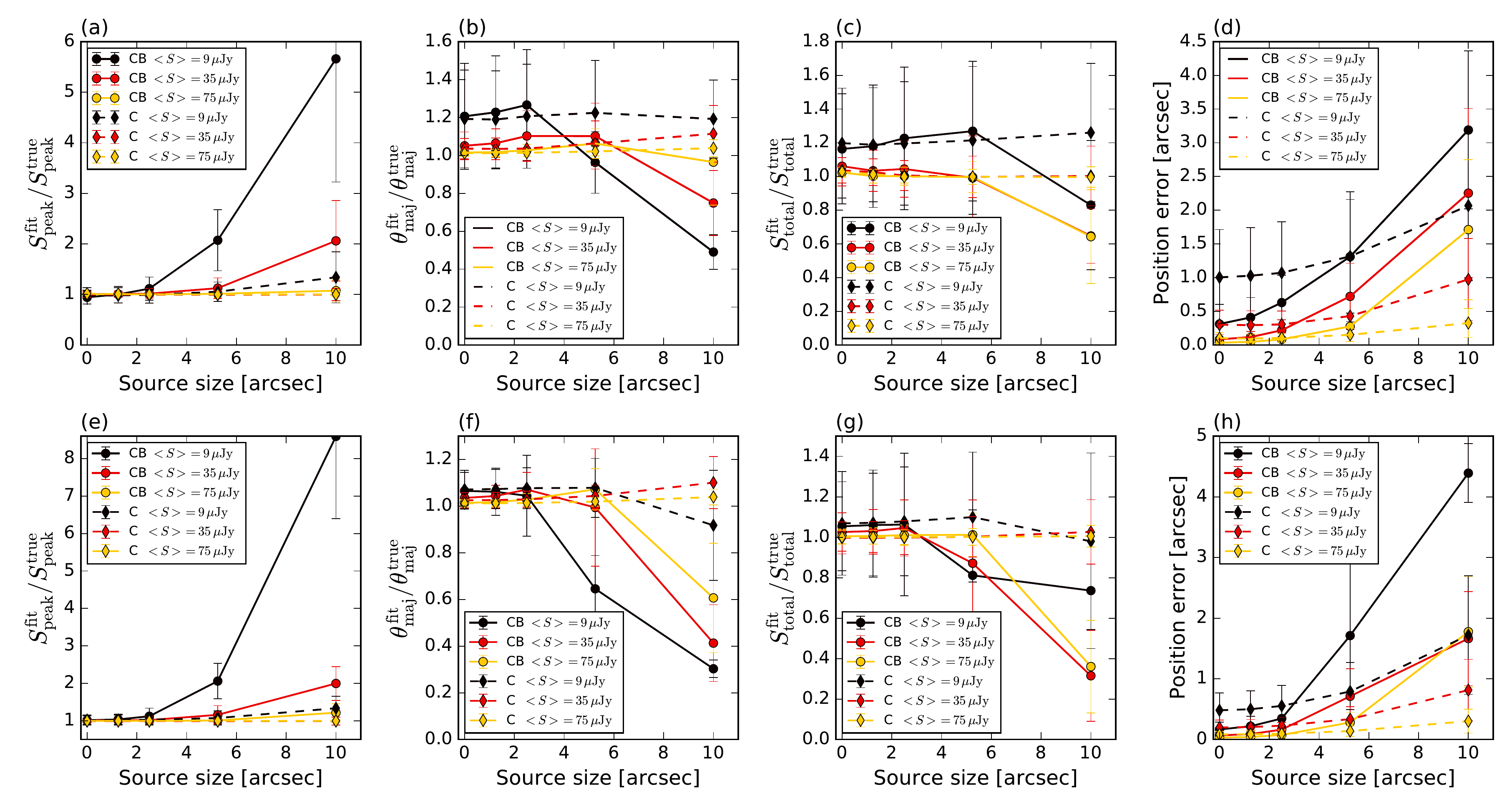}
\caption{Fit parameters compared to true parameters as a function of true source deconvolved size for matched sources for different flux bins. The top row is from {\Ob} and the bottom row is from \Ag. From left to right the panels are the ratio of fitted to true peak flux density, the ratio of fitted to true major axis, the ratio of fitted to true total flux density, and  the separation between the true and fitted source positions as a function of true source size. The solid lines, and circles, are for the CB image resolutions ($2.75\,$arcsec), while the dotted lines, and diamonds, are for the C image resolution ($8\,$arcsec). The flux bins are $3\le S \, \mu {\rm Jy} \le 15$ (black), $15\le S \, \mu {\rm Jy} \le 50$ (red), and $50\le S \, \mu {\rm Jy} \le 100$ (orange).}
\label{fig:sz_rats}
\end{figure*}

\section{Blending}
\label{sec:blend}
For each simulation realization there are approximately 1000 sources within the primary beam area that are ``detectable'' by the limits that we set. Among those 1000 for the worse resolution images, there are on the order of 50 to 100 sources that are fit as one source when there are in fact multiple (fainter) sources in or around the fitted source positions. In these ``blended'' cases the fitting software tends to fit large (12 to 20 arcsec convolved) sizes for one (though sometimes both) of the axes, and the total flux density for the one source is on average larger than either of the individuals it encompasses. This will have an impact on the source count of a catalogue; obviously decreasing the count of fainter sources and of the total number of sources, and misrepresenting the number of sources with total flux densities equal to the sum of the blended components. Source blending will also affect cross-matching with other wavelengths, i.e. if the reported position for the one fitted source does not match that of either of the component sources it could result in no matches or incorrect matches. Additionally, if we were to cross-match with another radio wavelength to measure spectral indices this would likely result in incorrect spectral-index estimates. 

In order to investigate this issue further we created a new set of simulated images. We generated 250 pairs of point sources (500 sources total), and picked total flux densities ranging over $4\le S_{\rm tot} \le 100 \, \mu$Jy, divided into 10 bins. The flux densities were assigned to ensure an even number of pairs with source flux densities from each bin (i.e. 25 pairs with $4 \le S_1/(\mu {\rm Jy}) \le 10$ and $4 \le S_2/(\mu {\rm Jy}) \le 10$, 25 pairs with $10 \le S_1/(\mu {\rm Jy}) \le 20$ and $4 \le S_2/(\mu {\rm Jy}) \le 10$, and so on). 

We generated random positions for one source in each pair, assuring that no pair was closer than $40 \,$arcsec to another pair. The position of the second source in each pair was then assigned to have a position with a distance to the first source of either 1.5, 3, 6, 8, or $12 \, $arcsec. These five images were then convolved with both the C image beam and the CB image beam, for a total of 10 images. Then beam-convolved Gaussian noise with $\sigma_{\rm C}=1.08 \, \mu$Jy beam$^{-1}$ and $\sigma_{\rm CB}=1.15 \, \mu$Jy beam$^{-1}$ was added to the images of the corresponding resolution. 

\begin{figure*}
\includegraphics[scale=.7,natwidth=9.5in,natheight=4.3in]{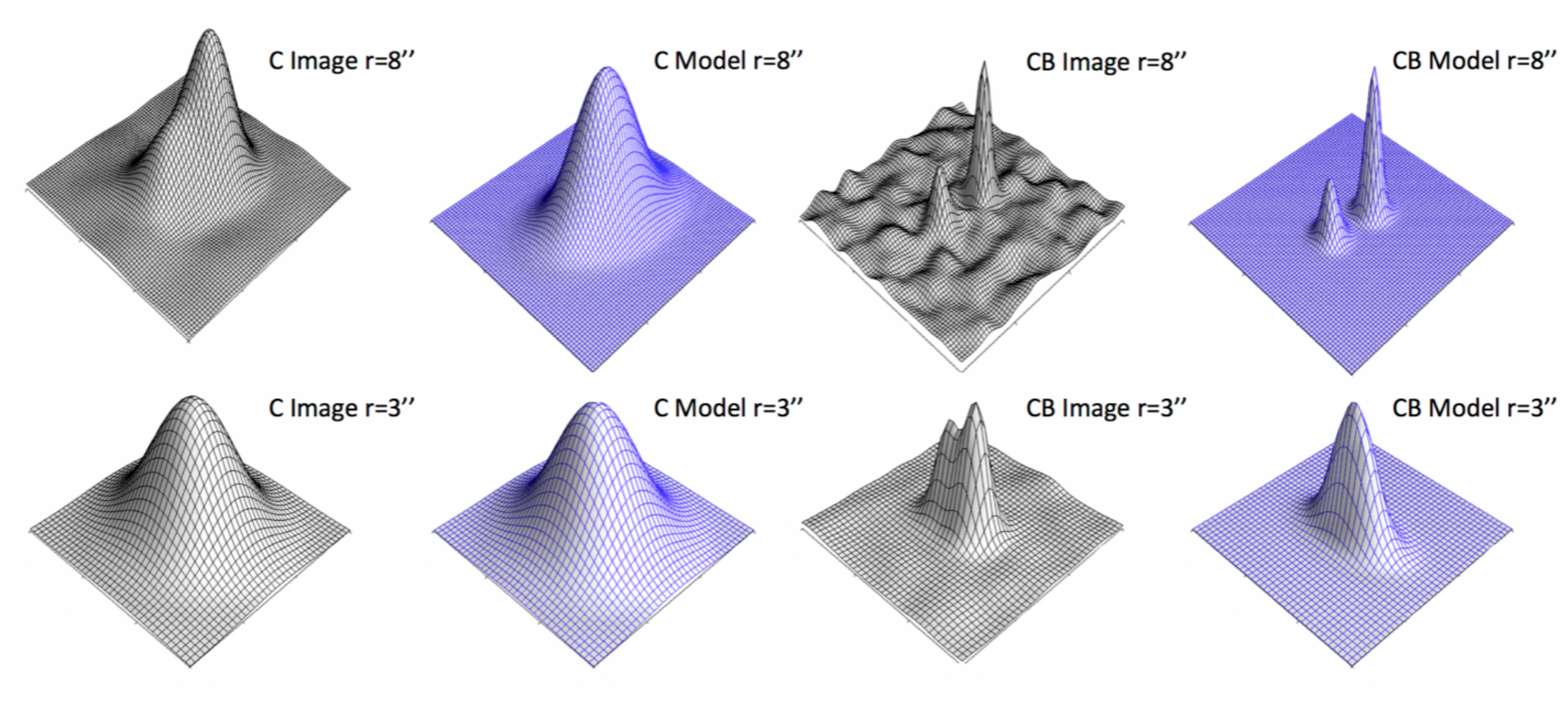}
\caption{Examples of simulated blended sources and model fits. Images are $25\,$arcsec on a side. The top row shows a pair of sources separated by $8\,$arcsec. The bottom row shows a pair of sources separated by $3\,$arcsec. From left to right the panels are the C image, C model fit, CB image, and CB model fit.  }
\label{fig:bl_3d}
\end{figure*}

These images were run through {\Ob} and {\Ag} to find and fit sources. Examples of one pair of sources, and their model fits at multiple separations, are shown in Fig.~\ref{fig:bl_3d}. The cross-matching was carried out by finding the closest fitted source for each input source, meaning that a fitted source could be matched with more than one input source. The fractions of input sources with a {\it unique} match in the fit results are presented in Fig.~\ref{fig:bl_frac}.

Figure~\ref{fig:bl_frac} shows that once the source separation is above \thetabeam, the fitting software is able to recognize two separate sources. For the cases where one source is fit for a pair of sources, the software, on average, conserves the total flux of the pair. Figure~\ref{fig:bl_ftots} shows the mean ratio of the sum of the total flux densities of a pair to the total flux density of the one fitted source, in bins of combined total flux density. Some cases have less lines plotted than others, this is because for those cases (software and resolution) there were no (or very few) blended sources at that particular separation. For example, the {\Ob} C data show lines extending to the $12\,$arcsec separation, whereas the {\Ag} CB data only show lines for the $1.5\,$arcsec and $3\,$arcsec separation, because at larger separations {\Ag} was easily able to distinguish the individual sources. The average ratios are all around unity, with a trend for deviating from this for sources with a larger separation. While it is reassuring that total flux is being conserved, this effect biases the source count, as well as the source size distribution. 

One might be able to do better by looking explicitly for blended sources. We briefly investigated the use of supervised machine-learning algorithms in identifying blended sources. We used the blended sources from this simulation and separate sources from the size simulation described above. Each source was given a classification, either a ``0" for blended or ``1" for unique. These classifications, along with the fit parameters of peak flux density, total flux density, major axis, minor axis, and ellipticity, were used to see if the machine-learning software could, using a subset of these as a training set, predict the correct classification for the remainder. For this test we used the \textsc{Scikit-Learn} python package \citep{Pedregosa11}.

The results showed that machine learning was able to predict the correct classification for $90\,$per cent of the test sources. This fraction could likely be improved using a larger training set and more fine-tuning of the algorithm. Using this technique one could then allow for those sources classified as blended to be refit, possibly using constraints on the fit parameters or using software capable of fitting a specific number of input Gaussian distributions. This type of identification is most useful when data are only available at one (poorer) resolution.  

Another approach to de-blending sources is to give users control over the number of Gaussian sources, or force a particular number of Gaussians, to be fit in a particular region. If multiple resolutions or data sets for the same data are available, and are comparably deep, this is a powerful aid in de-blending the sources. Essentially this is prior-based fitting. If there is prior knowledge that there are multiple sources (i.e. from other better resolution data in a region) allowing this information to be used in controlling the fits, either in number of sources and or source parameters, would decrease the problem of blended catalogue sources. {\Ag} includes an option for prior-based fitting by inputting a catalogue that {\Ag} then uses in the current fitting. The user has the option to vary all of the input parameters, only the flux and positions, or only the flux. 

As a test we tried this option using the images with sources separated by $6\,$arcsec. In the original fitting (no priors) the CB resolution was able to fit matches for all of the 500 input sources. The C resolution image only found the 250 pairs; it was unable to de-blend any of them. We used the CB {\Ag} output as the prior input on a new run to fit the C image. This resulted in finding 360 individual sources. The software did attempt to fit the other 140 individual sources found in the CB image; however the fitting failed, returning unreasonable values for the parameters.  

Clearly, this issue affects the worse resolution data much more than the better resolution data. The source count derived in \citet{Vernstrom13} predicts on average $0.15$ sources per $8\,$arcsec beam at $3\,$GHz for sources with $S>5\, \mu$Jy, and $0.7\,$sources per $10\,$arcsec beam at $1.4\,$GHz (the beam size and frequency of the ASKAP EMU survey), ignoring any clustering and additional blending from resolved sources. Thus, with the $8\,$arcsec beam (at $3\,$GHz), one can expect {\it at least} $1\,$per cent of sources to be closer than the beam size, and at least $5\,$per cent at $1.4\,$GHz with a $10\,$arcsec beam. The prior-based fitting and machine-learning identification help; however, there is still room and need for improvement. 

\begin{figure}
\includegraphics[scale=.36,natwidth=9in,natheight=9in]{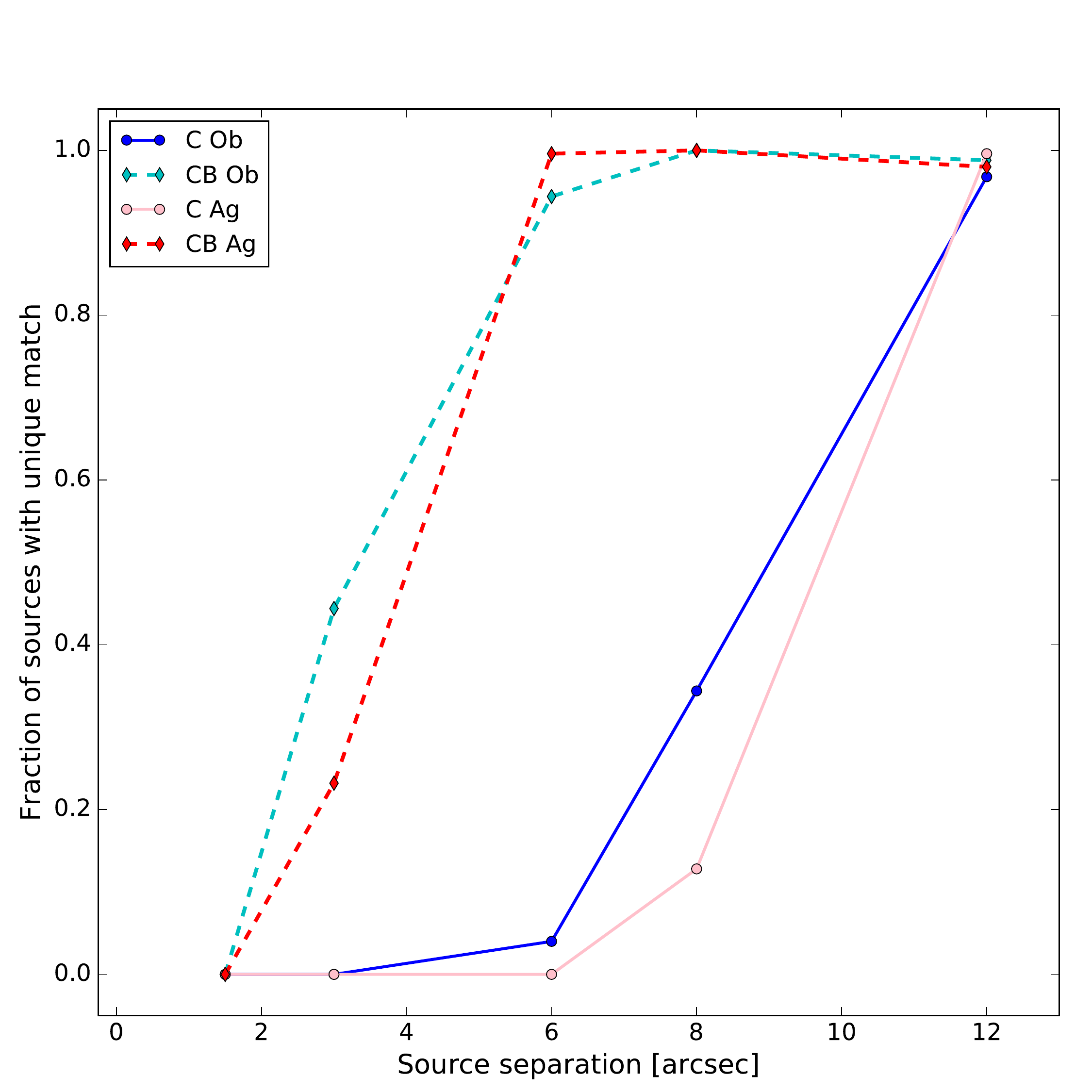}
\caption{Fraction of sources with a unique match as a function of input source separation. The solid dark blue line is the C image results from \Ob, the light blue dashed line is the CB image results from \Ob, the light red solid line is the C image results from \Ag, and the dark red dashed line is the results from the CB image from \Ag. }
\label{fig:bl_frac}
\end{figure}

\begin{figure}
\includegraphics[scale=.48,natwidth=7in,natheight=13.5in]{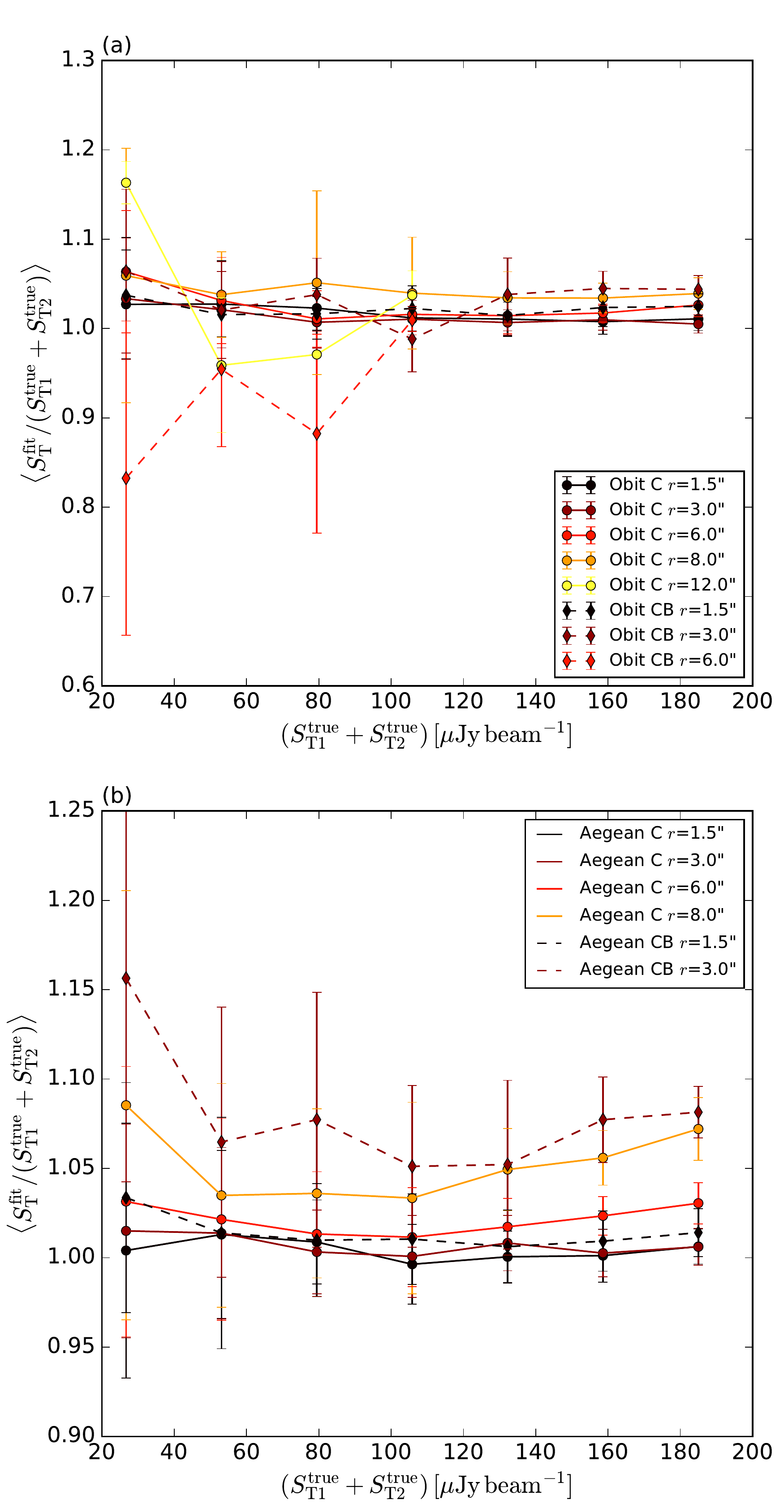}
\caption{Mean ratio of fitted total flux density to combined pair flux density in bins of the combined total flux density. The top panel shows the results from \Ob, while the bottom panel shows the {\Ag} results. The colours indicate the pair separation. Solid lines correspond to C data and dashed lines are for CB data.}
\label{fig:bl_ftots}
\end{figure}

\section{Discussion}
\label{sec:disc}

\subsection{Resolution comparison}
\label{sec:fres_comp}
There are certainly advantages and disadvantages when considering each resolution independently. Overall, the CB $2.75^{\arcs}$ beam yields smaller uncertainties on the fit parameters (except for the faintest-source axis sizes). Looking at the positional accuracy, the CB uncertainties on the RA and Dec estimates are 2--10 times smaller than for the worse resolution. For the peak and total flux densities and the axis sizes, the uncertainties for the CB data are 1--3 times smaller, except for the faintest fluxes, where the C resolution has slightly smaller uncertainties. However, at least with this particular noise setup, the false detection rate for the CB resolution near the detection limit is higher. 

The results for both resolutions suffer for sources with sizes roughly equal to or greater than the beam. Since it is believed that the majority of faint sources have sizes of less than approximately $1\,$arcsec \citep[e.g.][]{Muxlow07}, the better resolution data are the more affected. This also means that better resolution would likely be the poorer choice if one were interested in extended emission of sources. Not only will the better resolution miss many faint sources when the source size is greater than the beam size, but those it does find are not fit as accurately, with the peak flux densities being overestimated and source size, as well as total flux density, being underestimated. 

Despite appearing to be poor (by optical astronomy standards, for example), the C resolution of $8\,$arcsec still performs well (with this noise setup), finding more true sources and fewer non-real sources, and with reasonable fit uncertainties. The worse resolution has higher confusion noise (which can be good or bad, depending on the specific goal) as well as suffering from the blending issue. The amount of blending depends on the true level of source clustering, and while there are known algorithms and software for source de-blending, further studies are needed to make sure they perform as expected and perform optimally for the expected data with realistic levels of clustering.

When considering the question of which resolution to use, if given the choice of only one, the answer depends on the exact instrumental noise at each resolution; in our case the better resolution has slightly higher instrumental noise, although that is obviously not going to be the case when considering resolutions generally. The answer also depends on what science goals are hoped to be achieved. For example, \citet{Vernstrom13} required a higher confusion noise in order to use the {\it P(D)} analysis technique to estimate the source count of faint sources requiring a poorer resolution. However, if one is more concerned with cross-matching with other wavelengths and requires higher positional accuracy, then the better resolution is obviously the better choice. 

The best answer is clearly to obtain data at multiple configurations or resolutions, with roughly equal noise, if at all possible. Newer interferometric imaging techniques may also assist in some of the issues discussed above, such as using the multi-scale CLEAN algorithm \citep{Cornwell08}.

\subsection{Software comparison}
\label{sec:fit_comp}
In comparing the two different source-finding programs used, both performed comparably. However, there are some advantages and disadvantages to each, as well as areas where one appears to perform better than the other.

The uncertainties on the fit parameters are slightly smaller with the {\Ag} fits, but not significantly so. {\Ob} performs marginally better at de-blending objects, finding slightly more unique sources in the blending simulation at a given separation. Similarly, with the size simulation, at a given source size {\Ob} finds slightly more sources than \Ag, but in both cases tested the improvement is not significant. Both are comparable in their completeness with {\Ob} having on average 1.08 times higher completeness. {\Ob} shows a higher false detection rate, on average 1.75 times higher than \Ag. However, {\Ob} has a feature to constrain the fitting based on the estimated false detection rate. Using this feature on one simulated image showed an average decrease in the false detection rate of 20 to $50\,$per cent.   

{\Ob} has the option of more user control inputs and constraints. The user is allowed to define the fitting region, maximum number of islands, the number of search passes, flux cutoff, whether to fit multiple Gaussian distributions, and how many. There are also parameters the user can set to reject fits, including the maximum and minimum source size, the minimum peak flux density, peak position within fitting region, and more. {\Ob} can also automatically generate a residual image. 

{\Ag} has limited control over fit parameters, either in the fitting or fit acceptance. Also it allows for source sizes to be fit that are smaller than the beam size, creating an apparently imaginary deconvolved source size, and hence complicating the most appropriate values to report for size and total flux density. It also does not produce any residual image at the present time. However, {\Ag} uses a ``curvature" image, defined as the second derivative of the image using a discrete Laplacian operator (which can be saved as a FITS image along with the estimated background image and rms image). This curvature image is used to determine the number of components within an island of pixels and to produce a set of initial parameters and limits for a constrained fit of multiple elliptical Gaussians.  

While it may seem that {\Ob} has an advantage in allowing more user control, there is a danger of over-constraining the fitting and hence biasing the results. However, the differences in the {\Ob} and {\Ag} results are overall statistically insignificant, at least for our particular test data sets. 

\section{Conclusions}
\label{sec:cat_conc}

We have used multiple simulations to investigate the effects of image resolution, correlated noise, source size, source blending, and fitting accuracy for two different source extraction programs. For our pair of same-sky surveys at different resolutions these simulations yield estimates of the parameter uncertainties, as well as being useful for obtaining parameter corrections and completeness corrections. 

\begin{table*}
\centering
\caption{Summary of simulation fit statistics for $8\,$arcsec resolution using the {\Ag} software in bins of peak signal-to-noise ratio. Column 2, $f_{\rm matched}$, is the fraction of sources in the corresponding peak SNR bin that were found and had a match to the total number of true sources in that bin. Column 3, $f_{\rm false}$, is the fraction of sources in the corresponding peak SNR bin that were found and did not have a true match to the total number of sources found. }
\begin{tabular}{ccccccc}
\hline
\hline
$\langle$Peak SNR$\rangle$ & $f_{\rm matched}$& $f_{\rm false}$ & $\langle \Delta_{\rm position} \rangle$ & $\langle S_{\rm peak}^{\rm fit}/S_{\rm peak}^{\rm true} \rangle$& $\langle S_{\rm total}^{\rm fit}/S_{\rm total}^{\rm true} \rangle$ &  $\langle \theta_{\rm maj}^{\rm fit}/\theta_{\rm maj}^{\rm true} \rangle$ \\
&(percent)&(percent)&(arcsec)&&&\\
5&$79$ &$16$&$\pm0.85$&$1.17\pm0.18$ &$1.30\pm0.27$&$1.09\pm0.13$\\
15&$96$&$1.0$&$\pm0.57$&$1.06\pm0.14$&$1.18\pm0.21$&$1.10\pm0.11$\\
25&$97$&$0.0$&$\pm0.36$&$1.02\pm0.09$&$1.07\pm0.14$&$1.05\pm0.08$\\
50&$98$&$0.0$&$\pm0.26$&$1.01\pm0.05$&$1.03\pm0.09$&$1.03\pm0.06$\\
100&$99$&$0.0$&$\pm0.24$&$1.01\pm0.02$&$1.01\pm0.07$&$1.02\pm0.06$\\
\hline
\end{tabular}
\label{tab:suma_lr}
\end{table*}

\begin{table*}
\centering
\caption{Summary of simulation fit statistics for $2.75\,$arcsec resolution using the {\Ag} software in bins of peak signal-to-noise ratio. Column 2, $f_{\rm matched}$, is the fraction of sources in the corresponding peak SNR bin that were found and had a match to the total number of true sources in that bin. Column 3, $f_{\rm false}$, is the fraction of sources in the corresponding peak SNR bin that were found and did not have a true match to the total number of sources found.}
\begin{tabular}{ccccccc}
\hline
\hline
$\langle$Peak SNR$\rangle$ & $f_{\rm matched}$ & $f_{\rm false}$ & $\langle \Delta_{\rm position} \rangle$ & $\langle S_{\rm peak}^{\rm fit}/S_{\rm peak}^{\rm true} \rangle$& $\langle S_{\rm total}^{\rm fit}/S_{\rm total}^{\rm true} \rangle$ &  $\langle \theta_{\rm maj}^{\rm fit}/\theta_{\rm maj}^{\rm true} \rangle$ \\
&(percent)&(percent)&(arcsec)&&&\\
5&$70$&$17$ &$\pm0.44$&$1.15\pm0.20$ &$1.16\pm0.40$&$1.02\pm0.30$\\
15&$99$&$0.0$&$\pm0.18$&$1.04\pm0.12$&$1.09\pm0.19$&$1.05\pm0.10$\\
25&$99$&$0.0$&$\pm0.08$&$1.01\pm0.06$&$1.03\pm0.10$&$1.02\pm0.05$\\
50&$100$&$0.0$&$\pm0.05$&$1.00\pm0.05$&$1.01\pm0.05$&$1.02\pm0.04$\\
100&$100$&$0.0$&$\pm0.02$&$1.00\pm0.02$&$1.00\pm0.03$&$1.01\pm0.02$\\
\hline
\end{tabular}
\label{tab:suma_hr}
\end{table*}

\begin{table*}
\centering
\caption{Summary of simulation fit statistics for $8\,$arcsec resolution using the {\Ob} software in bins of peak signal-to-noise ratio. Column 2, $f_{\rm matched}$, is the fraction of sources in the corresponding peak SNR bin that were found and had a match to the total number of true sources in that bin. Column 3, $f_{\rm false}$, is the fraction of sources in the corresponding peak SNR bin that were found and did not have a true match to the total number of sources found.}
\begin{tabular}{ccccccc}
\hline
\hline
$\langle$Peak SNR$\rangle$ & $f_{\rm matched}$& $f_{\rm false}$ & $\langle \Delta_{\rm position} \rangle$ & $\langle S_{\rm peak}^{\rm fit}/S_{\rm peak}^{\rm true} \rangle$& $\langle S_{\rm total}^{\rm fit}/S_{\rm total}^{\rm true} \rangle$ &  $\langle \theta_{\rm maj}^{\rm fit}/\theta_{\rm maj}^{\rm true} \rangle$ \\
&(percent)&(percent)&(arcsec)&&&\\
5&$81$&$28$ &$\pm1.40$&$1.08\pm0.28$ &$1.52\pm0.28$&$1.30\pm0.19$\\
15&$95$&$2.0$&$\pm0.77$&$1.04\pm0.20$&$1.24\pm0.21$&$1.15\pm0.14$\\
25&$98$&$1.0$&$\pm0.43$&$1.02\pm0.13$&$1.10\pm0.14$&$1.06\pm0.09$\\
50&$98$&$0.8$&$\pm0.29$&$1.01\pm0.06$&$1.05\pm0.10$&$1.03\pm0.07$\\
100&$99$&$0.9$&$\pm0.26$&$1.01\pm0.02$&$1.03\pm0.12$&$1.02\pm0.05$\\
\hline
\end{tabular}
\label{tab:sumo_lr}
\end{table*}

\begin{table*}
\centering
\caption{Summary of simulation fit statistics for $2.75\,$arcsec resolution using the {\Ob} software in bins of peak signal-to-noise ratio. Column 2, $f_{\rm matched}$, is the fraction of sources in the corresponding peak SNR bin that were found and had a match to the total number of true sources in that bin. Column 3, $f_{\rm false}$, is the fraction of sources in the corresponding peak SNR bin that were found and did not have a true match to the total number of sources found.}
\begin{tabular}{ccccccc}
\hline
\hline
$\langle$Peak SNR$\rangle$ & $f_{\rm matched}$& $f_{\rm false}$ & $\langle \Delta_{\rm position} \rangle$ & $\langle S_{\rm peak}^{\rm fit}/S_{\rm peak}^{\rm true} \rangle$& $\langle S_{\rm total}^{\rm fit}/S_{\rm total}^{\rm true} \rangle$ &  $\langle \theta_{\rm maj}^{\rm fit}/\theta_{\rm maj}^{\rm true} \rangle$ \\
&(percent)&(percent)&(arcsec)&&&\\
5&$87$ &$60$&$\pm0.66$&$1.12\pm0.25$ &$1.38\pm0.32$&$1.19\pm0.24$\\
15&$99$&$1.0$&$\pm0.25$&$1.02\pm0.13$&$1.12\pm0.19$&$1.07\pm0.14$\\
25&$99$&$0.0$&$\pm0.10$&$1.01\pm0.07$&$1.05\pm0.10$&$1.03\pm0.07$\\
50&$99$&$0.0$&$\pm0.05$&$1.01\pm0.05$&$1.03\pm0.05$&$1.01\pm0.05$\\
100&$100$&$0.0$&$\pm0.02$&$1.01\pm0.02$&$1.02\pm0.02$&$1.00\pm0.02$\\
\hline
\end{tabular}
\label{tab:sumo_hr}
\end{table*}

When comparing the two catalogues of different resolution, the two approaches perform comparably well when it comes to finding the same sources and their parameters. Summaries of the fitting results for the different trials can be seen in Tables~\ref{tab:suma_lr}, \ref{tab:suma_hr}, \ref{tab:sumo_lr}, and \ref{tab:sumo_hr}. It is evident that sources near the SNR limit will have the least accurate fits. Our numbers show that for sources with peak flux densities $\le 10\sigma$ the fitted flux densities, as well as the sizes, are likely be overestimated by $10$ to $30\,$per cent, regardless of resolution or software. Using our simulated data to interpolate for parameter corrections based on the fitted peak flux density and fitted major axis, this bias can be reduced to roughly $3\,$per cent. 

From the simple set of simulations, set flux densities and no source sizes, but correlated noise, we found that adding the constraint of source size being larger than or equal to the beam size does create a bias for overestimated axis sizes, but also seems to remove the correlation between the peak flux density and axis sizes seen without the constraint. Using the more realistic setup simulations we found that the equation for parameter uncertainties from \citet{Condon97} for correlated noise works well except for the axis sizes. When computing uncertainties on the fitted sizes an additional dependence on the axis size and peak flux density is required; otherwise the uncertainties for faint small sizes are overestimated and those for bright large sizes are underestimated.

This method of using a large realistic set of simulated data to obtain corrections and uncertainties for real source fits should help to improve the fit accuracy, particularly for faint sources. The method could also be useful for future large surveys, but would require a larger set of simulations set up for the specific survey; these simulations would need to be performed in advance, for surveys such as EMU, which will carry out source extraction automatically as data are taken.

Using a different set of simulations we tested the effect of source size on fitting. We found that once the source size is roughly equal to the beam size the fraction of sources missed drops to $20\,$per cent and continues down to $75\,$per cent missed when the source is 3.5 times larger than the beam. For the resolved sources that are found, the results depend on the total flux density of the source, but the size is underestimated by 50 to $75\,$per cent when the source is 3.5 times larger (for flux densities between 3 to $50\,\mu$Jy), while the peak flux density is overestimated by 200 to $600\,$per cent for sources with flux densities between 3 to $50\,\mu$Jy. This results in the total flux density being underestimated by 60 to $80\,$per cent. This can severely affect source counts, as well as source characteristics such as spectral indices. The main way to solve this issue is by using multi-resolution data, though it may also be mitigated to some extent in the imaging stage with the use of multi-scale or tapered weighting imaging.

We also tested what effect source blending has on the fitting results. We found that once sources are separated by less than the beam size they are fit as one single source up to $95\,$per cent of the time. This also affects source count studies, as well as source characteristic studies, and creates difficulty in cross matching, as the position of the blended source is not usually aligned with the peak position of any of the component sources. Tests show that when using simulations as a training set for machine-learning algorithms, blended sources can be identified approximately $90\,$per cent of the time, and can then be refit with multiple Gaussians. We also tested one algorithm's prior-based fitting, using better resolution fitting results as a prior for the worse resolution fitting. This showed a marked improvement at fitting individual blended sources. 

The two programs that were compared, {\Ag} and \Ob, performed similarly. {\Ag} calculates the local rms, local background, and curvature across the image (which can also be saved as FITS images) and uses them in peak finding; tests have shown this rms estimation to be quite accurate. {\Ob} has the benefit of more user control over the fitting constraints. {\Ag} is a newer software package, which is still in development. In Paper \rom{2}, we present the catalogues for the actual VLA observations, for which all cataloguing uses the {\Ob} software. 

Although the effects discussed here are all well known, we have carefully quantified them, in the hope that this will be useful for future surveys. In the context of comparing future projects such as deep VLA (better resolution) versus EMU/MIGHTEE (worse resolution), we note that the better resolution data have the distinct advantage of providing more accurate positions and avoiding the problem of blending. On the other hand, the worse resolution data have the advantage of being able to find more of the extended sources that would have been missed or inaccurately fit in the better resolution \citep{Condon15}. Both resolution choices (as long as the noise is comparably deep) should find roughly the same number and the same sources and provide accurate estimates of their parameters, as long as careful attention is paid to the problem areas we have highlighted for each resolution. For these future deep surveys improvements will need to be made to existing imaging, source-finding, and fitting algorithms \citep[or new algorithms developed, e.g][]{Hollitt12,Frean14} in order to optimize the performance for faint ($\mu$Jy) sources and to carry out the procedure rapidly as the data pour in. It is in this faint regime that current software requires the most additional development to improve performance.

\section{Acknowledgments}
The Dunlap Institute is funded through an endowment established by the David Dunlap family and the University of Toronto. We acknowledge the support of the Natural Sciences and Engineering Research Council (NSERC) of Canada.  We thank the staff of the VLA, which is operated by the National Radio Astronomy Observatory (NRAO), a facility of the National Science Foundation operated by the Associated Universities, Inc.

\bibliographystyle{mnras}

\bsp

\label{lastpage}
\end{document}